\documentclass{ws-ijmpd}
\usepackage[super,compress]{cite}
\usepackage{bbding,enumitem}

\usepackage{pifont}%

\usepackage{subfigure}
\usepackage{multirow}
\usepackage{cancel}
\usepackage{array}
\usepackage{natbib}

\RequirePackage[colorlinks,citecolor=blue,urlcolor=blue,linkcolor=blue]{hyperref}

\usepackage{amsmath}
\usepackage{amssymb}
\usepackage{comment}
\usepackage{aas_macros}

\newcommand{\be}{\begin{equation}}
\newcommand{\ee}{\end{equation}}

\AtBeginDocument{}

\usepackage[normalem]{ulem} 

\begin{document}

\markboth{{\bf Kuo Liu and Siyuan Chen}}
{{\bf Pulsar Timing Array in the past decade}}

%
\catchline{}{}{}{}{}
%

\title{Pulsar Timing Array in the past decade}

\author{{\bf Kuo Liu}
}

\address{State Key Laboratory of Radio Astronomy and Technology, Shanghai Astronomical Observatory, CAS, 80 Nandan Road, Shanghai 200030, P. R. China\\
liukuo@shao.ac.cn}


\author{{\bf Siyuan Chen}
}

\address{State Key Laboratory of Radio Astronomy and Technology, Shanghai Astronomical Observatory, CAS, 80 Nandan Road, Shanghai 200030, P. R. China \\
siyuan.chen@shao.ac.cn }

\maketitle

\begin{abstract}
The past decade has been a transformative period for pulsar timing arrays (PTAs) and their search for nanohertz gravitational waves (GWs). This progress has been driven by collective advances in instrumentation for pulsar timing observations, increasingly sophisticated data-analysis techniques, and improved theoretical understanding of the origins of nanohertz GW signals. PTA sensitivity has steadily improved, leading first to progressively more stringent upper limits on the gravitational-wave background (GWB), and subsequently to the identification of a common red-noise process in pulsar timing data, the first hint of a GWB. In 2023, multiple PTA collaborations reported evidence for the Hellings--Downs correlation, widely regarded as the definitive signature of a GWB. These developments place PTAs on the threshold of a confident GW detection and the opening of a new low-frequency window on the GW Universe. In this article, we present an overview of PTA experiments, with particular emphasis on the rapid progress achieved during this pivotal period for PTA and nanohertz GW science.
\end{abstract}


\section{Introduction} \label{sec:intro}

Gravitational physics has been thrilling over the past decade, marked by a series of transformative breakthroughs that have reshaped the field. Foremost among these was the first direct detection of gravitational waves (GWs) in 2015 by the LIGO–Virgo–KAGRA Collaboration \citep{aaa+16a}, a discovery that confirmed a central prediction of general relativity and opened an entirely new observational window on the Universe. This advance was followed by the emergence of multi-messenger GW astronomy after the binary neutron star merger observed in 2017 \citep{aaa+17a}, which demonstrated the power of combining gravitational and electromagnetic observations. The first image of a black hole, released by the Event Horizon Telescope Collaboration in 2019 \citep{EHT:2019}, provided a direct view of spacetime in the immediate vicinity of an event horizon and offered striking visual evidence of strong-field gravity in action. In parallel, pulsar-based experiments have driven equally profound progress. The exceptional rotational stability of pulsars, together with the precision of modern pulsar timing techniques, has enabled tests of gravity in the strong-field regime to reach unprecedented precision \citep[e.g.,][]{agh+18,ksm+21}. Most notably, pulsar timing array (PTA) experiments have recently delivered a major breakthrough by reporting the first evidence for nanohertz GWs and the gravitational wave background (GWB) \citep{EPTA+2023c,ng15_gwb,pptadr3_gwb,cpta_dr1,meerkat_gwb}. This milestone heralds the opening of a new low-frequency window onto the gravitational-wave Universe, expanding GW astronomy to regimes inaccessible to ground-based detectors.

The milestone achieved by PTA experiments can be attributed to a range of significant developments across the field. On the observational side, substantial progress has been driven by the deployment of new, highly sensitive radio telescopes and the development of wideband signal reception techniques and high-rate digital signal processing systems. On the data-analysis side, increasingly sophisticated techniques and software packages have been developed, enabled in part by the growing affordability of high-performance computing, such as the continued improvement of CPU capabilities and the expanding use of GPU acceleration. In parallel, theoretical models for the origin of nanohertz GWs-—particularly predictions of the stochastic gravitational-wave background (GWB)—-have undergone important refinements. 

Most importantly, the PTA community itself has also grown and matured considerably. The International Pulsar Timing Array \citep[IPTA,][]{haa+10,vlh+16}, established in 2010, plays a central role in coordinating efforts among regional PTA collaborations. Its founding members include the European PTA \citep[EPTA,][]{kc13}, the North American Nanohertz Observatory for Gravitational Waves \citep[NANOGrav,][]{2009arXiv0909.1058J}, and the Parkes PTA \citep[PPTA,][]{hbb+09,mhb+13}. The collaboration has since expanded to include the Indian PTA \citep[InPTA,][]{tnr+22}, which joined in 2020, and the African Pulsar Timing (APT) community, which joined in 2025. In addition, the Chinese PTA \citep[CPTA,][]{cpta_dr1} and the MeerKAT PTA \citep[MPTA,][]{msb+23} participate as observing members and have established data-sharing agreements with the IPTA.

Coordinated global efforts within the PTA community have been a primary driver of the field’s breakthroughs over the past decade. The first generation of regional PTA datasets, together with the first IPTA data releases around 2013-–2016, placed increasingly stringent upper limits on the nanohertz GW signals \citep{src+13,ltm+15,abb+16,vlh+16}. Subsequent releases of expanded regional datasets culminated in the second IPTA data release in 2019 \citep{pdd+19}, which significantly improved overall sensitivity. A pivotal moment came in 2020–-2022, when both regional PTAs and the IPTA independently reported the detection of a common red-noise process in pulsar timing data -— the first indication of a GWB \citep{abb+20,ccg+21,gsr+21}. This progress led to another major milestone in June 2023, when the first coordinated, worldwide PTA data releases provided strong evidence for a nanohertz GWB \citep{EPTA+2023c,ng15_gwb,rzs+23,cpta_dr1}. This result was further supported by an independent confirmation from the MPTA in 2024 \citep{meerkat_gwb}. Complementary efforts have also extended PTA methodologies beyond radio observations. For example, a PTA-style analysis has been carried out using gamma-ray bright pulsars observed with the Fermi Space Telescope \citep[the Gamma-ray Pulsar Timing Array (GPTA),][]{2022Sci...376..521F}. At the moment, preparations are underway for a third IPTA data release, which is expected to combine the most recent datasets into the most sensitive global PTA dataset to date, incorporating timing data from more than 100 pulsars.

PTAs exploit the exceptional rotational stability and timing precision of millisecond pulsars (MSPs). Observations are typically conducted with the most sensitive radio telescopes at decimetre wavelengths, where MSPs are relatively bright. As a result, times-of-arrival (TOAs) of pulsar signals at Earth can be measured with precisions reaching tens of nanoseconds over timescales of decades. As gravitational waves (GWs) propagate through the Galaxy, they induce extremely small but correlated perturbations in the spacetime between the Earth and an ensemble of MSPs \citep{saz78,det79}. The hallmark signature of a stochastic GW background is a distinctive quadrupolar pattern of spatial correlations in the timing signals of different pulsars, known as the Hellings–Downs (HD) correlation \citep{hd83}. PTAs are sensitive to GWs in the nanohertz to microhertz frequency range, set by the observational baseline and observing cadence. This frequency band is complementary to the higher-frequency regimes probed by ground-based detectors such as LIGO and by future space-based detectors such as LISA.

The primary GW signal targeted in the nanohertz band is the GWB, a stochastic signal formed by the superposition of numerous individual GW sources. The most promising contributor is the population of inspiralling supermassive black hole binaries (SMBHBs) in the local Universe \citep{1980Natur.287..307B}. SMBHBs are a natural outcome of galaxy mergers, in which the central supermassive black holes of the progenitor galaxies are driven toward the center of the merged system through dynamical friction and other interactions. Once the binary separation becomes sufficiently small, GW emission is expected to dominate the inspiral. The cumulative signal from the entire SMBHB population is therefore predicted to form a stochastic background \citep{1995ApJ...446..543R}. In addition, particularly massive or nearby SMBHBs may be individually resolvable in frequency and sky location, introducing anisotropies in the background. The identification of such individual sources would open new opportunities for multi-messenger GW astronomy by enabling direct comparisons with electromagnetic observations.

Beyond SMBHBs, a nanohertz GWB may also arise from cosmological processes or exotic physics in the early Universe. These scenarios can produce backgrounds with spectral shapes and spatial-correlation signatures distinct from those expected from SMBHB populations \citep[e.g.,][]{2023ApJ...951L..11A,2024A&A...685A..94E}. The standard HD correlation is derived within the framework of general relativity (GR), which assumes massless gravitons, propagation at the speed of light, and only two GW polarization modes. Any statistically significant deviation from the HD correlation could therefore indicate new gravitational physics beyond GR. A confirmed detection of such deviations would have profound implications for our understanding of gravity and cosmic evolution. In this sense, PTAs serve not only as GW detectors but also as precision laboratories for testing the fundamental laws of nature.

The aim of this article is to provide an overview of PTA experiments, with particular emphasis on the key developments in the field over the past decade. The review is organized into six sections. Section~\ref{sec:timing} summarizes the fundamentals of pulsar timing, covering observations, time-of-arrival measurements, and timing and noise models. Section~\ref{sec:PTA} describes how PTAs function as gravitational-wave detectors. In Section~\ref{sec:result}, we present the major results achieved by the PTA community over the last decade. Section~\ref{sec:origin} discusses possible astrophysical, cosmological, and exotic GW sources. Finally, Section~\ref{sec:conclu} concludes with a summary and an outlook for future developments. Table~\ref{tab:symbols} lists the symbols used throughout this article.

\begin{table}[h!]
    \tbl{Common symbols used in this review.}
    {
    \scriptsize
    \renewcommand{\arraystretch}{1.4} 
    \centering
    \begin{tabular}{|c|c|}
    \hline
    SYMBOL & DESCRIPTION \\ \hline \hline
    $P(t)$ & Waveform of observed profile \\ \hline
    $T(t)$ & Waveform of profile template \\ \hline
    DM & Dispersion measure  \\ \hline
    S/N & Signal-to-noise ratio \\ \hline
    $t_{\rm psr}$ & Pulsar proper time \\ \hline
    $t_{\rm obs}$ & Observed times of arrival (TOAs) \\ \hline
    $\mathbf{r}$ & Pulsar Timing Residual \\ \hline
    $\mathbf{s}$ & Delays induced by the signals in the model \\ \hline
    $\nu$ & Spin frequency \\ \hline
    $\nu_{\rm r}$ & Radio frequency \\ \hline
    $f$ & Fourier frequency \\ \hline
    $f_{\rm r}$ & Reference frequency (commonly set at 1/yr) \\ \hline
    $\sigma_{\rm rn}$ & TOA uncertainty by radiometer noise \\ \hline
    $\sigma_{\rm J}$ & TOA uncertainty by jitter noise \\ \hline
    $\sigma_{\rm scint}$ & TOA uncertainty by diffractive scintillation \\ \hline
    $n_e$ & Free electron density \\ \hline 
    $d$ & Distance of pulsar to Earth \\ \hline
    $c$ & Speed of light \\ \hline
    $e$ & Electron charge \\ \hline
    $m_e$ & Electron mass \\ \hline
    $\tau_{\rm scat}$ & Scattering timescale \\ \hline  
    $S(f)$ & Power spectral density (PSD) \\ \hline
    $A$ & Amplitude of a power law PSD model\\ \hline
    $\gamma$ & Spectral index of a power law PSD model \\ \hline    
    $h_c(f)$ & Characteristic strain of the GWB \\ \hline
    $\Omega_{\rm GWB}(f)$ & Cosmological Density of the GWB \\ \hline
    $\zeta$ & Inter-pulsar Angular Separation \\ \hline
    $\Gamma(\zeta)$ & Overlap Reduction Function (ORF) \\ \hline
    $C$ & Covariance matrix \\ \hline
    $\hat{n}$ & Unit vector---Earth to GW source \\ \hline
    $F_A(\hat{n})$ & Antenna pattern functions \\ \hline
    \end{tabular}
    \label{tab:symbols}
    }
\end{table}

\section{Pulsar Timing}  \label{sec:timing}

Pulsar timing is the foundational technique for constructing a PTA ``detector'' for nano-hertz GWs. Pulsars are commonly understood to be rapidly rotating, magnetised neutron stars, with a typical diameter of about 20\,km, a mass of approximately 1.4\,M$_{\odot}$, and a surface magnetic field strength of order $10^{12}$\,G. According to the standard stellar evolutionary scenario, a neutron star is one of the three possible end states following the death of a star that evolves from the main-sequence phase into the giant branch \citep{st83}. A pulsar can be viewed as a highly magnetised, rotating sphere surrounded by a co-rotating magnetosphere. Its radio beam is generated by the acceleration of charged particles by strong electric fields in a region above the magnetic pole \citep[e.g.][]{rs75}. In each rotation, when the emission beam sweeps past the Earth —- like a celestial lighthouse —- a pulse is observed.

\begin{figure}
    \includegraphics[width=1\linewidth]{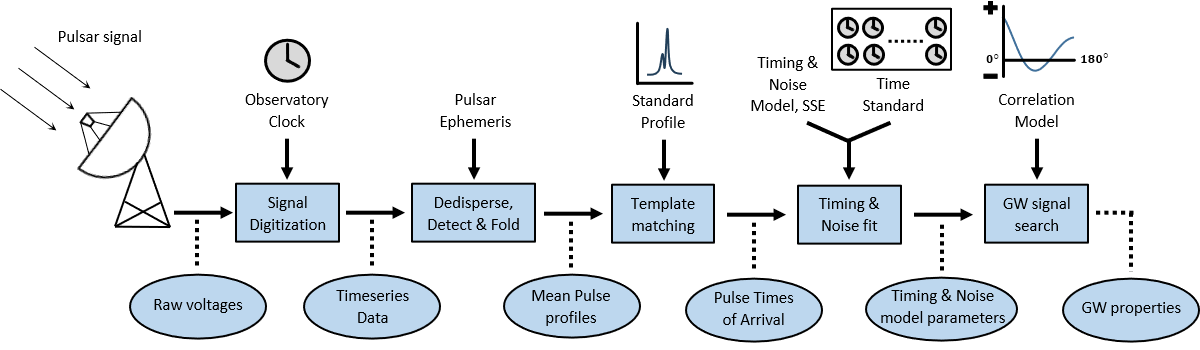}
    \caption{Sketch showing the overall data processing chain and products of each stage, from a typical contemporary PTA experiment.\label{fig:pta_process}}
\end{figure}

The technique of pulsar timing exploits pulsars as celestial clocks. Decades of observations have demonstrated that pulsars with rotational period $P<30$\,ms, commonly referred to as millisecond pulsars (MSPs), have long-term timing stability and thus are highly sensitive probes of tiny perturbations in spacetime \citep[e.g.,][]{EPTA+2023a}. Pulsar timing is conducted by measuring the pulse time-of-arrivals (TOAs) at Earth and maintaining the counts of pulses across a long timeline. A sketch summarizing major data processing stages in a PTA experiment can be found in Figure~\ref{fig:pta_process}. 

\subsection{Data acquisition of pulsar timing observation} \label{ssec:timing obs}

An astronomical radio signal from a pulsar can be described as a plane-propagating, transverse electromagnetic wave with two orthogonal polarisation states. After being focused by the radio antenna, the signal is collected by a receiver equipped with two probes sensitive to the two polarisations, converted into complex voltages, and amplified by a cryogenically cooled low-noise amplifier. To minimise attenuation during transmission through cables and render the signal suitable for later recording, it is typically down-converted from the observational radio frequency (RF) to an intermediate frequency (IF). The signal is then passed to a pulsar backend, which samples and digitises the voltage data at the Nyquist rate. In many systems, the backend applies the polyphase filterbank technique to the time-series data to divide the full bandwidth into a number of frequency sub-bands. 

Next, the data are passed to computing devices which process the signal to remove the dispersion of the pulsar radiation caused by interstellar medium (see Section~\ref{ssec:noise}), and to form intensity with full-polarisation information. The data are then folded with respect to the pulsar period, using an input pulsar ephemeris, which contains information about the pulsar’s rotational period, period derivative, binary parameters, and related quantities (see Section~\ref{ssec:timing model}). This procedure produces archival data products consisting of mean pulse profiles with a specified integration length and frequency resolution.

In the past, at the typical pulsar timing observing frequency of decimeter wavelengths, receiver bandwidths were usually of order a few hundred megahertz. To conduct pulsar timing observations over a broad frequency range, it was therefore necessary to observe the same pulsars in multiple frequency bands sequentially \citep[e.g.,][]{dcl+16}. Over the past decade, substantial development in receiver technology has enabled receivers to cover a continuous ultra-broad frequency range, for example 700 MHz-–4\,GHz at the Parkes Radio Telescope \citep{hmd+20} and 1.3--6.0\,GHz at the Effelsberg Radio Telescope\footnote{\url{https://eff100mwiki.mpifr-bonn.mpg.de/doku.php?id=information_for_astronomers:rx:p170mm}}. For interferometric arrays, the use of subarray observing modes can also provide simultaneous coverage of multiple frequency bands, as demonstrated with the Giant Metrewave Radio Telescope \citep{pkc+19}. The application of these newly developed receiver systems has shown greatly improved efficiency of pulsar timing observations.

Legacy pulsar backends developed approximately two decades ago typically delivered bandwidths of only a few hundred megahertz \citep[e.g.,][]{dcl+16,vlh+16}. Many of these systems were not capable of performing coherent dedispersion, leaving the pulsar signal smeared within each frequency channel. Owing to limitation in memory capacity and data-processing speed, the data were usually sampled with low-bit resolution (e.g., 1-bit or 2-bit), which introduced artefacts and distorts the recorded pulsar signal \citep[e.g.,][]{ja98,lvk+11}. The new generation of pulsar recording systems employs high-speed analogue-to-digital (A/D) converters and field-programmable gate array (FPGA) technology\footnote{See e.g., \url{http://casper.berkeley.edu/}, \url{https://www.realdigital.org/hardware/rfsoc-4x2}}, which enable high-bit (e.g., 8-bit) real-time data sampling and filterbanking over significantly wider bandwidth, up to a few gigahertz \citep[e.g.,][]{ctg+13,pbb+15,lkg+16,hmd+20}. These system are further equipped with a cluster of CPU/GPU computing nodes each of which process a fraction of the bandwidth to perform real-time coherent dedispersion, intensity detection and folding of the pulsar signal. 

In addition, recent observing systems have begun to place the digitisation hardware directly near the receiver feed, for example in the focal cabin \citep{hmd+20,mbb+25}. In this configuration, the analogue signal is converted to digital form immediately after being collected and is then transmitted via optical fibre to backend processors for further reduction. This approach improves signal fidelity by minimising analogue signal paths, avoiding signal degradation and interference pickup that can occur over long analogue cable runs.

\subsection{TOA creation} \label{ssec:TOA}

The mean pulse-profile data are usually subjected to further reduction steps, such as the removal of radio-frequency interference, polarisation calibration, and additional time and frequency averaging \citep[e.g.,][]{van03a,lvk+11}. Once these procedures are completed, the next step is to estimate the time-of-arrival (TOA) of the pulse profile. 

TOA measurements are typically fulfilled by a template-matching technique, in which a template profile-intended to represent the intrinsic shape of the pulsar signal—is cross-correlated with the observed mean pulse profile. The template-matching method is based on the assumption that the observed pulse profile can be described by
\begin{equation}
P(t)=a+bT(t-\Delta\tau)+n(t), \label{eq:model-T}
\end{equation}
where $P(t)$ represents the observed profile, $T(t)$ denotes the template, $a$ is an arbitrary baseline offset, $b$ is a scaling factor, $n(t)$ is the additive noise on the observed pulse profile caused by radiometer noise in the observing system, and $\Delta\tau$ is the phase offset between the profile and the template. The TOA is then calculated from the fitted phase shift $\Delta\tau$ together with the epoch of the observation. The template is usually generated from a high–signal-to-noise (S/N) pulse profile with radiometer noise removed \citep{kwj+94,hvm04}.

To find the best estimate of the phase shift between the observed profile and template, a least-squares fit is usually performed in the Fourier domain, which enables the determination of the phase offset less than one phase bin. After applying a discrete Fourier Transform to both the profile and the template, Eq.~(\ref{eq:model-T}) can be written as \citep{tay92}:
\begin{equation}
P_{k}e^{i\theta_{k}}=aN+bT_{k}e^{i(\phi_{k}+k\Delta\tau)}+n(k),
~~k=0,\cdots,~(N-1) \label{eq:model-F}
\end{equation}
where $N$ is the number of bins in the profile, $P_k$ and $T_k$ are the Fourier amplitudes of the profile and template, respectively, $\theta_k$ and $\phi_k$ are their corresponding phases in the $k$th bin, and $n(k)$ represents the noise contribution in the Fourier domain. Estimates of the phase shift $\Delta\tau$ and the scaling factor $b$ are obtained by minimising the goodness-of-fit statistic
\begin{equation}
\chi^{2}(b,\Delta\tau)=\sum^{N-1}_{k=1}\left|\frac{P_{k}-bT_{k}e^{i(\phi_{k}-\theta_{k}+k\Delta\tau)}}{\sigma_{k}}\right|^{2},
\end{equation}
where $\sigma_{k}$ is the root-mean-square (rms) intensity of the noise at frequency $k$. 

The traditional approach described above utilises one-dimensional profile data. In reality, however, pulse profiles are multidimensional, with dependency on frequency, time, and polarisation. The inclusion of polarisation information in template matching has been shown to improve TOA precision when the pulsar signal is significantly polarised \citep{van04a,ovd+13}. In recent years, wide-band observing systems have become available \citep{pkc+19,hmd+20}, delivering profile data with broad frequency coverage. This provides a powerful means to measure the DM at each observing epoch accurately. Traditionally, it can be achieved by dividing the data into a number of sub-bands, measuring TOAs from each individually, and deriving both the DM and the TOA at infinite frequency from these measurements. This procedure is usually referred to as ``narrow-band timing''. Alternatively, an epoch DM and TOA can be obtained by performing template matching directly on frequency-resolved profile data. If the frequency resolution of the profile data is retained, the template-matching model can be written as \citep{ldc+14,pdr14}
\begin{equation}
P(f,t) = a(f) + b(f)T\left(f,t - \Delta\tau - \mathcal{D}\times \Delta{\rm DM}/f^2 \right) + n(f,t),
\label{eq:mod_t}
\end{equation}
The corresponding goodness-of-fit statistic is
\begin{equation}
\chi^2(b_i,\Delta\tau,\Delta{\rm DM})=\sum^{N_{\rm
c}}_{j=1}\sum^{N-1}_{k=1}\left|\frac{P_{j,k}-b_jT_{j,k}e^{i(\phi_{j,k}-\theta_{j,k}+k\Delta\tau)}}{\sigma_{j,k}}\right|^2,
\label{eq:chisqr_DM}
\end{equation}
where $\Delta{\rm DM}$ is the difference in DM between the observed profile and the template. This quantity can either be fixed or fitted for during template matching. If the latter approach is adopted, the derived TOAs include a correction for DM differences between the profiles and the template, and are usually referred to as “wide-band” TOAs. In this case, the timing data do not need to be explicitly modelled for DM variations in the subsequent noise analysis (see Section~\ref{ssec:noise}).

Frequency-resolved template can be created by first independently generating templates for each frequency band and then concatenate them to form a two-dimensional template \citep[e.g.][]{lgi+20,kmj+21}. Alternatively, the two-dimensional (phase–-frequency) portrait of the profile data can be decomposed into principal components, and a selected subset can be used to construct a noise-free representation of the profile while capturing its frequency evolution \citep{pen19}. This method can be further extended to incorporate polarisation information, yielding a three-dimensional template profile \citep{czw+25}.

To date, wide-band timing technique has been applied in only a handful of studies \citep[e.g.][]{lgi+20,aab+21,kmj+21}, all of which have produced timing results consistent with those obtained from narrow-band timing analyses. Because wide-band timing yields significantly fewer TOAs than the narrow-band approach, it substantially reduces the volume of timing data. Therefore, it is anticipated that future adoption of wide-band timing will greatly accelerate PTA data analysis.

\subsection{The deterministic timing model} \label{ssec:timing model}

Pulsars are precision clocks in their own inertial reference frames. The cumulative number of pulsar rotations ($\mathcal{N}$) since a reference epoch ($t_0$) can be approximated by a Taylor expansion:
\begin{equation}
\mathcal{N}=\mathcal{N}_0+\nu_0(t_{\rm psr}-t_0)+\frac{1}{2}\dot{\nu}_0(t_{\rm psr}-t_0)^2+\frac{1}{6}\ddot{\nu}_0(t_{\rm psr}-t_0)^3+\cdots,
\end{equation}
where $t_{\rm psr}$ is the proper time measured at the pulsar, $\nu_0$ is the spin frequency at $t_0$. 

In practice, TOAs are measured with respect to a local time standard maintained by an atomic clock at the observatory. This observatory time is usually converted to Coordinated Universal Time (UTC) using the Global Positioning System (GPS) and subsequently related to Terrestrial Time (TT). Furthermore, they need to be transformed from those in the observatory reference frame to the pulsar proper time ($t_{\rm psr}$) to allow for rotation counting. This transformation must account for several effects, including the orbital motion of the Earth around the Sun, the relative motion between the Solar System barycentre (SSB) and the pulsar, and the orbital motion of the pulsar itself if it resides in a binary system. This sequence of time transformations is collectively referred to as the pulsar timing model. At the top level, the pulsar timing model can be written as
\begin{equation}
t_{\rm psr}=t_{\rm obs} - \Delta_{\odot} - \Delta_{\rm IS} - \Delta_{\rm bnry},
\end{equation}
where $\Delta_{\odot}$ includes delays associated with the coordinate transformation from the observatory to the SSB frame and propagation delays of the pulsar signal within the Solar System; $\Delta_{\rm IS}$ denotes the transformation to the barycentre of the pulsar system and the propagation delays arising from the relative motion between the two barycentres and from the passage of the signal through the interstellar medium (ISM); and $\Delta_{\rm bnry}$ includes the transformation to the pulsar’s inertial frame, vacuum delays due to the pulsar’s orbital motion, and propagation delays caused by the passage of the signal through the gravitational field of the companion star. The commonly used pulsar timing parameters can be categorised as given below \citep{ehm06}:
\begin{itemize} { 
    \item {\bf Spin parameters:} These describe the rotational behaviour of the pulsar. They include the spin frequency and its time derivatives, as well as sudden changes in the pulsar spin, usually referred to as glitches.
    \item {\bf Astrometry parameters:} These describe the position and motion of the pulsar relative to Earth. They include the two-dimensional position of the pulsar on the sky, their secular time variations (proper motion) and the annual parallax.
    \item {\bf Dispersion parameters:} These describe the time delays caused by the ionised plasma along the propagation path of the pulsar signal to the Earth. They include the dispersion measure (DM) and its time derivatives, as well as the amplitude of the solar-wind–induced interplanetary dispersion delay. 
    \item {\bf Keplerian parameters:} For pulsars in binary systems, these include the five parameters that describe the Keplerian orbit: Longitude of periastron, projected semi-major axis, orbital eccentricity, orbital period, and epoch of periastron passage. For nearly circular orbits, a different parameterisation is usually used to mitigate the degeneracy among the orbital parameters \citep{lcw+01}. 
    \item {\bf Post-Keplerian (PK) parameters:} The PK parameters include secular variations of the Keplerian parameters, which can be induced by relativistic, kinematic or environmental effects \citep[e.g.][]{lk04}. They also characterise relativistic time-delay effects, such as time-dilation and gravitational redshift (Einstein delay), and propagation delay of the pulsar signal due to spacetime curvature induced by the companion (Shapiro delay). }
\end{itemize}

Computing the time delay terms also requires accurate, time-dependent model of the positions, velocities, and related physical properties of bodies in the Solar System (including the Sun, the Earth--Moon system, the other major planets and Pluto, and a number of asteriods). Usually, this information is included in a {\it solar system ephemeris} as a given input for the timing analysis \citep[e.g.,][]{pfw+21}, while some of these properties can also be modelled with PTA data \citep{chm+10,cgl+18}. 

To obtain the best estimates of the timing parameters, a least-squares fit is performed to find the minimum of the $\chi^2$ statistic,
\begin{equation}
\chi^2=\sum_i\left(\frac{\mathcal{R}_i}{\sigma_i}\right)^2,
\end{equation}
where $\mathcal{R}_i$ is the difference between the $i$th observed TOA and the value predicted by the timing model -- usually called the timing residual -- and $\sigma_i$ is the measurement uncertainty of the $i$th TOA. The uncertainties can be computed from the covariance matrix of the paramter estimates.

Timing analysis is widely performed using the \textsc{tempo} and \textsc{tempo2} software packages \citep[e.g.][]{hem06}, as well as the \textsc{pint} software toolkit which has been developed in recent years \citep{lrd+21}. In addition to the aforementioned packages that employs a frequentist approach, timing analysis can also be carried out within a Bayesian framework to estimate timing parameters and derive their credible intervals. This capability is provided by the \textsc{temponest} software package which operates as a plugin for \textsc{tempo2} \citep{lah+14}, and, more recently, by the \textsc{Vela.jl} software package \citep{sus25}. 

\begin{figure}
\hspace*{-1.8cm}
    \includegraphics[scale=0.35]{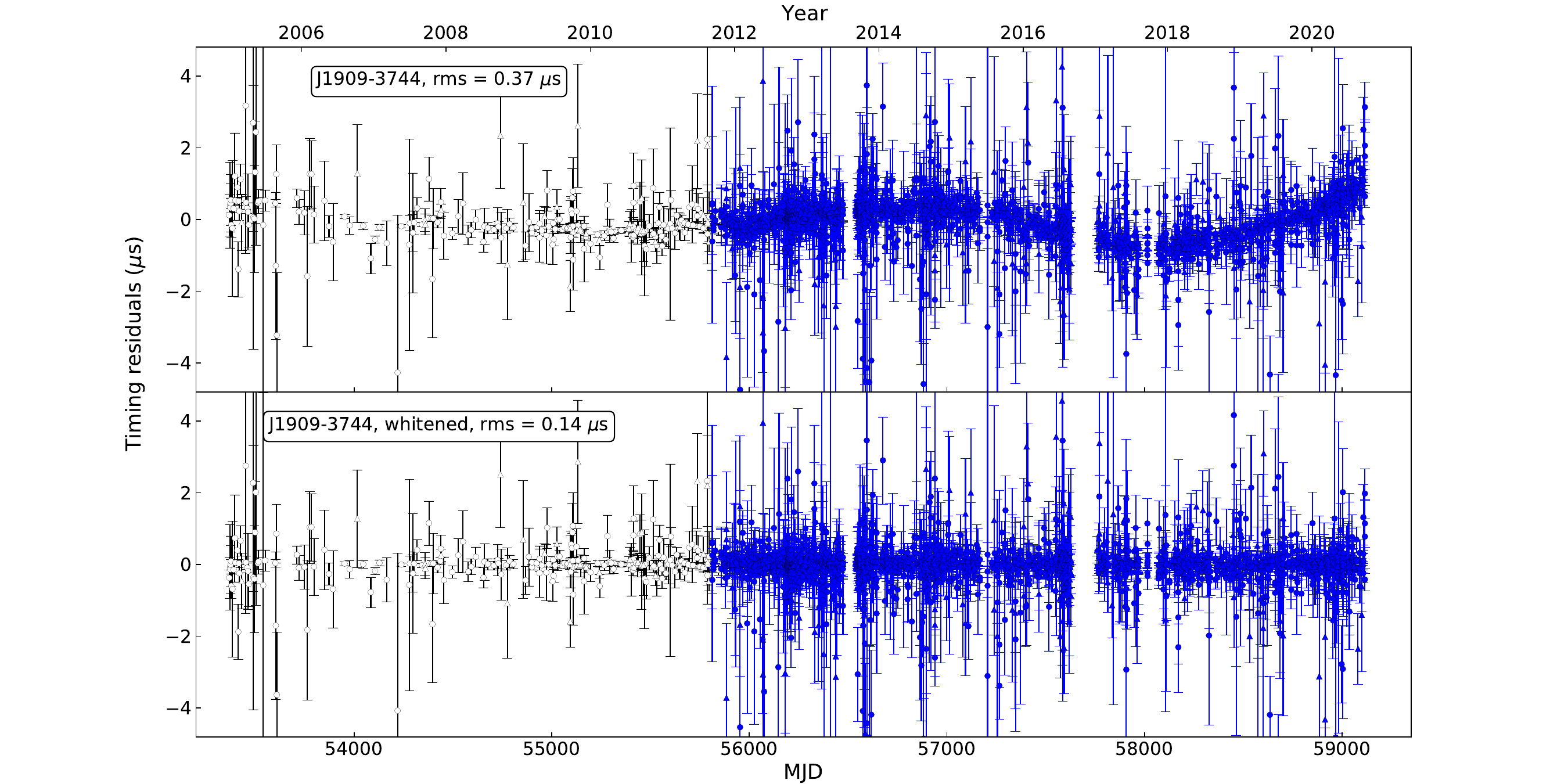}
    \caption{Timing residuals of PSR J1909$-$3744 from approximately 16-yr timing observations, before and after subtracting the red noise present in the dataset. The figure was produced using EPTA Data Release 2 data from \cite{EPTA+2023a}. \label{fig:res}}
\end{figure}

It is worth noting, that in addition to the deterministic timing model, pulsar timing data also contain stochastic features that must be modelled (see Figure~\ref{fig:res}). These features are usually referred to as timing noise (see Section~\ref{ssec:noise}). If not properly taken into account, they can bias the inferred timing parameters \citep[e.g.,][]{mcc13,lpk+25}. It is anticipated that the global, joint modelling of timing and noise parameters can often deliver more robust estimates of the timing parameters. This capability is now provided by most modern pulsar timing software packages that has been developed in recent years, including \textsc{temponest} \citep{lah+14}, \textsc{enterprise} \citep{ska+24}, and \textsc{Vela.jl} \citep{sus25}.

\subsection{Noise characterization in pulsar timing} \label{ssec:noise}

Understanding the noise budget and correctly measuring the noise components in a dataset are essential for both timing analyses of individual pulsars and gravitational-wave (GW) searches using multiple pulsars as a PTA, since imperfect noise modelling can bias the inferred properties of other signals. At the level of a single pulsar, there are two main types of stochastic noise components in the data: white noise and red noise. The red-noise component can be further categorised into chromatic and achromatic red noise, which are dependent on and independent of the radio frequency of the timing observation, respectively. The noise components associated with different pulsars are, in general, independent and uncorrelated. At the PTA level, however, additional noise sources may be present that are correlated between different pulsars.

\subsubsection{White noise sources}

{\bf Radiometer noise}

Ideally, the uncertainty of a TOA stems mainly from the additive noise on the observed pulse profile induced by the radiometer noise of the observing system. This can be written in the form of \citep{dr83}:
\begin{equation}
\sigma^2_{\rm rn}=\frac{\Delta t}{N (\rm S/N_{1})^2\int [\hat{T}'(t)]^{2}dt}, \label{eq:sigma_rn}
\end{equation}
where $\rm S/N_{1}$ is the averaged single-pulse peak signal-to-noise ratio (S/N), $N$ is the number of pulses integrated, $\hat{T}(t)$ is the peak-normalised profile waveform and $\Delta t$ is the sampling interval. In principle, this is also the TOA uncertainty derived from the template matching method discussed in Section~\ref{ssec:TOA}. 

The level of radiometer noise depends directly on the system sensitivity, which is largely determined by the collecting area of the telescope and the bandwidth of the recorded data. Since 2019, usage of the Five-hundred-metre Aperture Spherical Telescope (FAST), the largest single-dish radio telescope on Earth, has significantly enhanced the sensitivity of pulsar timing observations \citep{jyg+19}. In parallel, interferometric techniques have been employed to coherently combine signals from a large number of individual dishes. This can be achieved for arrays of dishes within relatively short baselines (typically $\lesssim 1$ km), such as the MeerKAT Radio Telescope \citep{jm16}, the Very Large Array\footnote{\url{https://science.nrao.edu/facilities/vlba/docs/manuals/oss2016A/prop-prep/phsd-vla}}
, the Atacama Large Millimeter/submillimeter Array \citep{mcd+18,lyw+19}, and, in the near future, the Square Kilometre Array (SKA). For widely separated telescopes, very long baseline interferometry (VLBI) can also be used to increase the effective collecting area. For example, the Large European Array for Pulsars (LEAP) project coherently sums the signals from the five 100-m–class radio telescopes in Europe. This configuration delivers an equivalent collecting area corresponding to a single dish of up to 194 m in diameter \citep{bjk+16}, rivaling that of the initial phase of the SKA currently under construction \citep{jkb25}.

{\bf Jitter noise}

The radio pulse emission from individual rotations of a pulsar exhibits a pulse-to-pulse variability. This phenomenon has been observed in canonical pulsars for decades, and, more recently, in MSPs \citep[e.g.,][]{lkl+12,ovb+14,lkl+15,lbj+16,pbs+21,lab+22}. This variability can be a stochastic process which introduces random phase shifts of integrated pulse profiles and, consequently, in their TOA measurements. This resultant jitter noise can be quantified as:
\begin{equation}
    \sigma_{\rm J}^{2}=f_J^{2}\frac{\int \hat{T}(t)t^{2}dt}{N\int \hat{T}(t)dt}, \label{eq:sigma_j}
\end{equation}
where $N$ is the number of pulses, $f_{J}$ is the Gaussian probability width of the single pulse phase in units of the pulse width. On small frequency scales, jitter noise is fully correlated across the observing band and has a uniform amplitude \citep[e.g.,][]{lkl+12}. However, it is likely not completely broadband, as in some cases jitter noise has been shown to decorrelate over bandwidths of order GHz \citep[e.g.,][]{sod+14,ksr+24} and to have its amplitude vary with frequency \citep[e.g.,][]{sod+14,lma+19,pbs+21}.

A comparison of Equations~(\ref{eq:sigma_rn}) and (\ref{eq:sigma_j}) shows that the relative significance of radiometer and jitter noise depends primarily on observational sensitivity and the intrinsic variability of the pulsar. While extending the integration time can mitigate jitter noise, in this case it imposes a fundamental limit on timing precision and thus the overall efficiency of timing observations. Nonetheless, additional dedicated techniques may be applied to the pulse profile data to further reduce jitter noise, such as adaptive fitting to compensate for profile variations \citep[e.g.,][]{ovh+11,nma+23} and the clustering of individual pulses \citep[e.g.,][]{ovb+14,ker15,lbj+16,slm25}.

{\bf Scintillation noise}

The ISM between a pulsar and Earth scatters and scintillates the radio signal from the pulsar. Scattering temporally broadens the pulsar signal, shifting its centroid and, consequently, the measured time of arrival (TOA). Stochastic variation in this pulse broadening arises from the finite number of scintles in the scatter-broadened image, a phenomenon known as the finite scintle effect \citep{cs10}. This variability introduces white noise into the timing data, which can be approximated as:
\begin{equation}
\sigma_{\rm scint}^{2}\simeq\frac{t^2_{\rm d}}{N_{\rm scint}}, \label{eq:sigma_scint}
\end{equation}
where $t_{\rm d}$ is the pulse broadening timescale, and $N_{\rm scint}$ is the number of scintles contained in the observation used to form the pulse profile. While typically negligible compared to radiometer and jitter noise at the canonical timing frequency ($\sim$1\,GHz), this noise contribution becomes increasingly prominent at lower observing frequencies \citep{lcc+16}.

{\bf Systematics}

Timing data can be affected by various sources of systematic error. These include instrumental or processing artifacts that distort the observed pulsar signal, thereby introducing a systematic bias into its TOA measurement. Common examples include radio frequency interference (RFI), imperfect polarization calibration, low-bit sampling, filterbank spillover and so forth \citep{ja98,van03a,lvk+11}. An additional source of systematic white noise can arise from the template-matching procedure used to measure TOAs, particularly when there is a significant mismatch between the template shape and the intrinsic pulse profile \citep{lvk+11}.

{\bf Modelling of white noise}

As a result of development in noise modelling techniques in recent years, white noise is now usually modeled with three parameters: \texttt{EFAC} ($E_{\rm f}$), \texttt{EQUAD} ($E_{\rm q}$)) and \texttt{ECORR} \citep[e.g.,][]{abb+15,EPTA+2023b}. \texttt{EFAC} is a multiplicative factor that scales the TOA uncertainty obtained from template matching, to account for its potential underestimation or overestimation. \texttt{EQUAD} is a noise term added in quadrature to represent additional white noise beyond the radiometer limit, as discussed previously. Together, these parameters modify the effective TOA uncertainty as follows:
\begin{equation}
    \sigma^2=E^2_{\rm q}+E^2_{\rm f}\sigma^2_{\rm rn}.
    \label{eq:whitenoise}
\end{equation}
These parameters are typically defined per observing system, usually corresponding to a unique receiver/backend combination. The \texttt{ECORR} parameter represents additional white noise that is fully correlated across all frequency channels within a single observing epoch, but uncorrelated between epochs. It captures effects such as pulse jitter and other band-correlated instrumental systematics. In the residual covariance matrix, the contribution from \texttt{ECORR} for observing system $k$ is expressed as:
\begin{equation}
    \mathcal{J}_{ij,k}=J^2_{k}U_{ij}, \label{eq:ECORR}
\end{equation}
where $i$, $j$ denote TOA indices across all epochs, $J^2_{k}$ is the variance of the epoch-correlated noise for system $k$, $U_{ij}$ is a block-diagonal matrix with entries of 1 for TOAs from the same epoch and 0 otherwise.

\subsubsection{Chromatic red noise sources}

Chromatic red noise in pulsar timing data arises primarily from variations in the propagation of the radio signal through the ISM along the line of sight from the pulsar to the Earth. These variations are driven by both the relative motion between the pulsar, Earth, and the intervening ISM, and by the intrinsic turbulence of the ISM plasma itself. Red-noise processes are typically modeled as a stationary, stochastic process with a power-law spectrum:
\begin{equation}
S(f)=\frac{A(\nu_{\rm r})^2}{C_{\rm 0}}\left(\frac{f}{f_{\rm r}}\right)^{-\gamma},
\label{eq:rednoise}
\end{equation}
where $S(f)$ is the spectral density at Fourier frequency $f$, $A(\nu_{\rm r})$ is the amplitude (dependent on the observing radio frequency $\nu$), $\gamma$ is the spectral index, and $f_{\rm r}$ is a reference frequency, conventionally set to 1\,yr$^{-1}$. The constant $C_{\rm 0}$ is a scaling factor that depends on the specific implementation of the noise model \citep[e.g.,][]{EPTA+2023b}. The spectrum includes a low-frequency cutoff, determined by the inverse of the total time span of the dataset. The amplitude of a chromatic red-noise component scales with the observing radio frequency as 
\begin{equation}
    A(\nu_{\rm r})\propto\nu_{\rm r}^{-\alpha}, 
\end{equation}
where $\alpha$ is the chromatic index. For modelling DM variations, $\alpha=2$. For scattering variations, the chromatic index depends on the scattering screen model and can be set, for example, to $\alpha=4, 4.4$ \citep{bcc+04}. Alternatively, it can be treated as a free parameter in the analysis.

{\bf DM variation}

When radio waves propagate through the ISM, the plasma disperses the radiation as a result of the frequency-dependent refractive index, causing lower frequency signals to arrive later than higher frequency ones. The resulting dispersive time delay can be written as \citep[e.g.][]{lk05}
\begin{equation}
\Delta t=\mathcal{D}\times\frac{\rm DM}{\nu_{\rm r}^2}. \label{eq:DM delay}
\end{equation}
where ${\rm DM}$ is defined as the integrated number density of free electrons $n_e$ along the line of sight,
\begin{equation}
{\rm DM}=\int_0^d n_e dl,
\end{equation}
with $d$ denoting the distance from the pulsar to Earth. The dispersion constant $\mathcal{D}$ is given by
\begin{equation}
\mathcal {D}=\frac{e^2}{2\pi m_e c}.
\end{equation}
where $e$ and $m_e$ are the electron charge and mass, respectively, and $c$ is the speed of light.

As is evident from Eq.~(\ref{eq:DM delay}), DM variations introduce frequency-dependent delays in the measured TOAs of the pulsars. Precise measurement of the DM and its temporal variation, therefore, requires sufficiently broad multi-frequency coverage \citep[e.g.][]{ldc+14}. The same requirement applies to the other chromatic effects in pulsar timing data, which are discussed below. Traditionally, DM variations were modelled using low-order polynomials or piecewise windowing functions \citep[e.g.][]{yhc+07,kcs+13,NG15+a}. Over the past decade, however, it has become increasingly common to model DM variations as a stochastic process \citep[e.g.][]{EPTA+2023b,rzs+23,ng156_noise}.

{\bf Scattering variation}

Interstellar scattering arises from multi-path propagation of the pulsar signal caused by electron-density fluctuations in the ionised ISM. If the medium along the line of sight is inhomogeneous, the pulsar wavefront is distorted during propagation. Thus, radiation emitted simultaneously into a small solid angle at the pulsar can reach the Earth along multiple paths, but at slightly different times. This effect broadens the observed pulse profile and results in an effective shift in the measured TOA. The observed pulse profile can be expressed as the convolution of the intrinsic pulse profile with a pulse-broadening function,
\begin{equation}
P_{\rm obs}(t) = P_{\rm int}(t) \ast \mathcal{B}(t).
\end{equation}
Assuming scattering is caused by a thin slab screen, the pulse-broadening function takes the form of \citep{sch68,wil72}
\begin{equation}
\mathcal{B}(t) = \frac{1}{\tau_{\rm scat}} e^{-t/\tau_{\rm scat}} U(t),
\end{equation}
where $U(t)$ is the unit step function. Assuming variations in the electron density show a single typical size $a$, the scattering timescale $\tau_{\rm scat}$ is given by
\begin{equation}
\tau_{\rm scat}=\frac{e^4d^2}{4\pi^2m^2_e a}\Delta n^2_e \nu_{\rm r}^{-4}.
\end{equation}
In practice, the inhomogeneity in the ISM can be considerably more complex and can lead to different power-law scalings of $\tau_{\rm scat}$ with observing radio frequency. For example, a Kolmogorov turbulence spectrum predicts $\tau_{\rm scat} \propto \nu_{\rm r}^{-4.4}$ \citep{lkm+01}. As with dispersion, scattering varies with time due to the relative motion of the pulsar, Earth, and the intervening ISM, as well as intrinsic turbulence within the plasma. These temporal variations introduce a chromatic red-noise component into pulsar timing data, and can be distinguished from DM noise through wide-band observations. Recent analyses have identified significant scattering noise in a number of PTA pulsars \citep[][]{rzs+23,EPTA+2023b,mac+23}.

{\bf Solar wind}

The solar wind contains free electrons flows and introduces an interplanetary dispersion delay in the radio emission from pulsars \citep[e.g.,][]{imn+98}. Traditionally, in pulsar timing the solar-wind effect is described using a spherically symmetric model, parameterised by the number density of free electrons at a heliocentric distance of 1~AU \citep[$n_{\rm sw}$; e.g.,][]{ehm06}. The corresponding dispersive delay, $\Delta_{\rm sol}$, can be written as \citep[e.g.,][]{ehm06}:
\begin{eqnarray} \label{eq:sol_delay}
    \Delta_{\rm sol}= \frac{{\rm DM}_{\odot}}{{\rm D}_{\rm 0}} \cdot \frac{1}{\nu_{\rm r}^2}
    &=&\frac{n_{\rm sw}(1\,\rm AU)^2\theta(E)}{{\rm D}_{\rm 0}|\vec{r}(E)|\sin \theta(E)} \cdot\frac{1}{\nu_{\rm r}^2},
\end{eqnarray}
where ${\rm D}_{\rm 0}=2.410\times10^{-16} \rm cm^{-3}~pc$, $\vec{r}(E)$ is the Earth–Sun distance, and $\theta(E)$ is the pulsar-Sun-observatory angle. The sole free parameter in this model, $n_{\rm sw}$, represents the mean electron density of the solar wind at 1~AU. Owing to the relative motion of the pulsar, Sun, and Earth, the solar-wind dispersion delay exhibits a significant annual modulation.

Early PTA analyses applied this static solar-wind model to correct for the interplanetary dispersion delay \citep[e.g.,][]{ehm06}. However, this approach neglects temporal variability in the solar-wind density (effectively variations in $n_{\rm sw}$), and has been shown to be inadequate for noise modelling, leaving clear residual signatures, particularly near solar conjunction \citep{tvs+19}. Such unmodelled solar-wind variability can bias other noise and timing parameters or mimic low-frequency signals of astrophysical interest \citep{lcc+16,mca+19,lpk+25}. The solar-wind density is known to vary on long timescales associated with the $\sim$11-year solar cycle, and on shorter timescales especially induced by the fast solar wind \citep[e.g.,][]{sbt+13,poi+14,stz20,tsb+21}. To account for this variability, recent noise analyses-analogous to treatments of DM and scattering variations-have adopted more flexible models for the solar-wind contribution. In particular, Gaussian process models have been used to describe the solar-wind amplitude as a time-variable stochastic process \citep{hsm+22,nkt+24}. This approach improves substantially over the standard static correction and has revealed strong correlations between solar-wind variability and pulsar ecliptic latitude \citep{sct+24}.

\subsubsection{Achromatic red noise sources}

{\bf Spin noise}

Canonical pulsars are known to exhibit rotational irregularities that cause red noise in the timing data that is independent of observing frequency \citep[e.g.,][]{gro75b,lhk+10}. Although MSPs exhibit much more stable rotations, they are also likely to possess some level of intrinsic instability \citep[e.g.,][]{sc10,EPTA+2023b}. For instance, this is supported by the detection of glitches in a small number of MSPs so far, including some that are used in PTA experiments \citep{cb04,mjs+16,bkc+26}. Consequently, spin noise must also be considered in MSP timing analyses.

Indeed, single-pulsar noise studies have shown that a large fraction of MSPs exhibit achromatic red noise \citep[e.g.,][]{EPTA+2023b,rzs+23,ng156_noise}. However, at the level of individual pulsars, it is difficult to distinguish spin noise from GW signals, as both manifest as achromatic red-noise processes. In multi-pulsar analyses, spin noise is in principle uncorrelated between pulsars, but it can still lead to false indications of a common noise process \citep{zhs+22}. Therefore, the use of population-level models for spin noise has recently been proposed as a way to obtain unbiased constraints on GW signals \citep{gts+22,van24}.

{\bf Profile variation}

As noted in Section~\ref{ssec:TOA}, standard pulsar timing analysis fundamentally assumes that the observed pulse profiles are intrinsically stable. Violations of this assumption -- where the profile shape varies over time -- inevitably introduce systematic errors into the measured TOAs. While profile variations have been well-documented in canonical pulsars for decades \citep[e.g.,][]{ls04}, they have only recently been identified in a small number of MSPs. This variability can manifest as sudden, significant distortions in the pulse profile \citep[e.g.,][]{ssj+21,jcc+24}, as gradual evolution over time scales of hours \citep[e.g.,][]{lkl+15,slk+16}, or as mode-switching between distinct profile states \citep[e.g.,][]{msb+22,nma+23}. Such variability introduces achromatic red noise into the timing residuals. Advanced techniques that directly model the evolving profile concurrently with the timing solution offer a promising method to isolate this noise component from other astrophysical signals in the data \citep[e.g.,][]{lkd+17,msb+22,nma+23}.

\subsubsection{Correlated noise sources}

Most timing noise is largely uncorrelated between different pulsars. However, certain noise components are common to all pulsars within a PTA. They can introduce distinctive spatial correlations analogous to the quadrupolar signature induced by a GWB \citep{hd83,thk+16}. If not carefully characterized and modelled, these common-noise correlations can significantly bias or even produce false claims of GWB detection. For instance, a monopolar correlation--where all pulsars are affected identically--can arise from errors in the terrestrial time standard to which all observatory clocks are referred to \citep{EPTA+2023b}. A dipolar correlation, conversely, is typically induced by errors in the SSE used to transfer observatory times to the Solar System Barycenter \citep[e.g., DE440;][]{pfw+21}, as positional errors impart a sky-dependent timing delay \citep{abb+18b}. The potential covariance between SSE errors and a GWB signal depends on the precision, time span, and number of pulsars in the dataset. Notably, analyses with recent PTA datasets have become increasingly robust showing diminishing effects from the specific SSE model chosen \citep{abb+20,ccg+21,gsr+21}. Furthermore, as all lines-of-sight to PTA pulsars pass through the solar wind, unmodeled variations in solar-wind density may induce correlations that mimic a GWB signal \citep{thk+16}. In particular, pulsars at very low ecliptic latitudes ($\lesssim 5^\circ$) show DM-variation trends consistent with the slow solar wind, constituting a potential common red-noise process \citep{sct+24}.


\subsubsection{Noise modelling approaches}

\begin{figure}
    \includegraphics[scale=0.33]{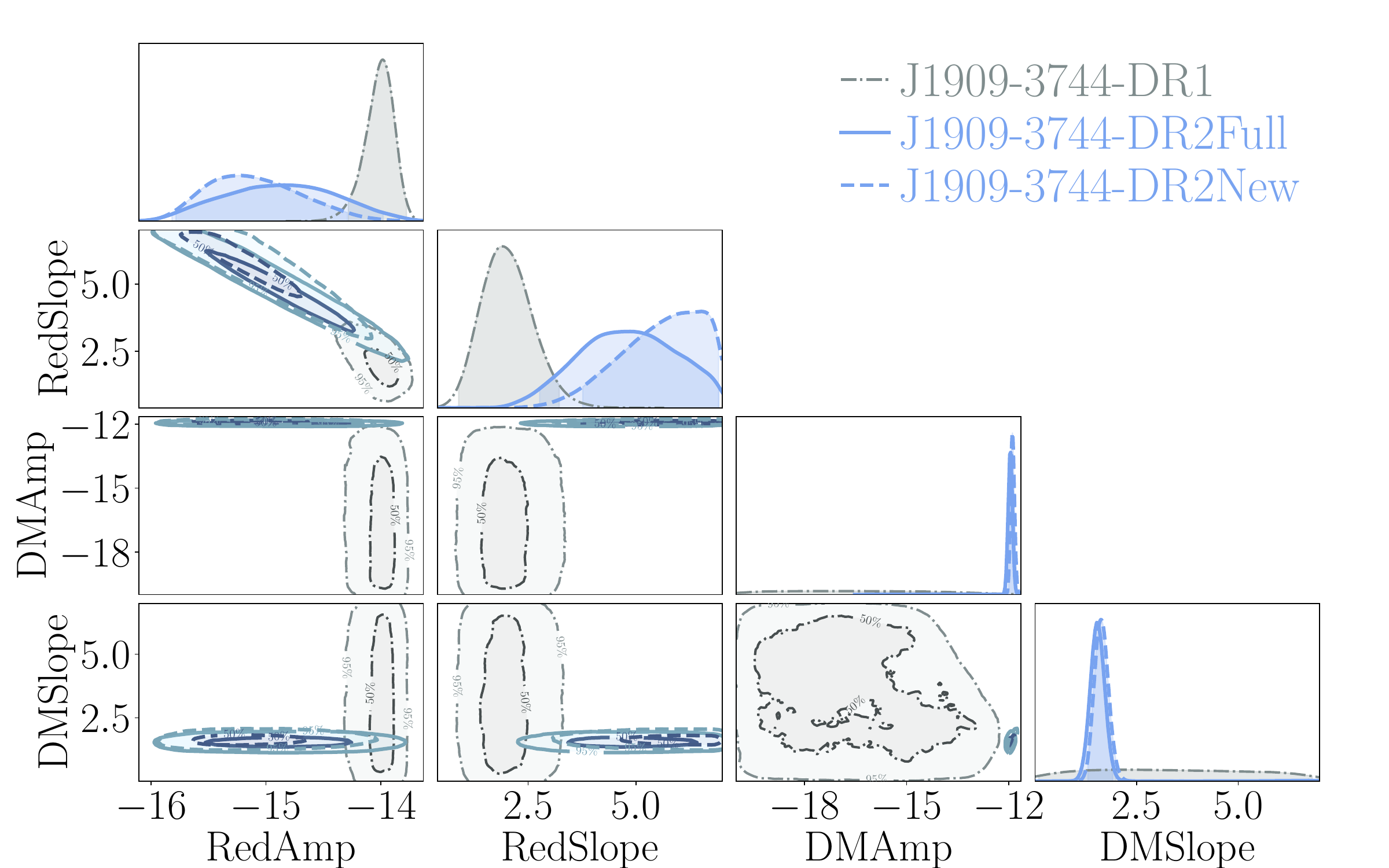}
    \caption{Marginalized posterior distributions for the noise parameters (spectral amplitude and index) of both DM and achromatic red noise for PSR~J1909$-$3744. These measurements were derived from three different versions of the EPTA dataset, where the EPTA DR2 provides significantly improved frequency coverage in the timing data compared to DR1 \citep[see][for details]{EPTA+2023a}. The figure is reproduced from \cite{EPTA+2023b}. \label{fig:noise_corner}}
\end{figure}

As discussed above, a precise separation of chromatic and achromatic red noise requires timing data spanning a wide range of radio frequencies to break the degeneracy among noise components (see also Figure~\ref{fig:noise_corner} as an example). The impact of limited frequency coverage in PTA datasets, and the potential biases it can introduce in inferred noise components and GW properties, has been studied extensively in recent years \citep{lmc+18,lgi+20,ict+24}. Typically, achieving multi-frequency coverage required conducting consecutive timing observations of each pulsar at multiple frequency bands. However, as discussed in Section~\ref{ssec:timing obs}, the development of new generation pulsar observing systems has enabled simultaneous wideband frequency coverage \citep[e.g.,][]{pkc+19,hmd+20,swk+21}. The application of these systems greatly improves the efficiency of PTA observations and can significantly mitigate chromatic noise arising from DM, scattering, and solar-wind variations.

In addition to the canonical spectral-fitting approach used to whiten timing residuals \citep[e.g.,][]{chc+11,lbj+14}, Bayesian inference techniques have become widely adopted for modelling pulsar timing noise and for selecting the noise model that best describes the data. For a sequence of measured timing residuals $\mathbf{r}$, the parameters $\mathbf{\theta}_i$ of a chosen model $\mathbf{M}_i$ are treated as random variables whose probability density functions (the posteriors) are given by Bayes' theorem \citep[e.g.,][]{EPTA+2023b}:
\begin{equation}
    p( \mathbf{\theta}_i | \mathbf{r}, \mathbf{M}_i) = \frac{p( \mathbf{r} | \mathbf{\theta}, \mathbf{M}_i) \ p( \mathbf{\theta}_i | \mathbf{M}_i)}{p( \mathbf{r} | \mathbf{M}_i)},
    \label{BayesTh}
\end{equation}
where $p( \mathbf{r} | \mathbf{\theta}, \mathbf{M}_i)$, $p( \mathbf{\theta}_i | \mathbf{M}_i)$ and $p( \mathbf{r} | \mathbf{M}_i)$ denote the likelihood, the priors of the parameters and the model evidence, respectively. A Gaussian likelihood is typically assumed, in which deterministic signals are absorbed into the timing residuals and stochastic processes are modeled through the time-domain covariance matrix $C$ \citep{vv15,EPTA+2023b}. 

Customizing a pulsar timing noise model primarily includes selecting the noise components necessary to describe the data and determining the number of Fourier modes used to represent each red-noise process. The selection of noise components is typically guided by Bayesian model selection, which provides a statistical framework for comparing competing models. One can evaluate the probability ratio between two candidate models, $\mathbf{M}_1$ and $\mathbf{M}_2$, known as the Bayes factor:
\begin{equation}
    \mathcal{B}^{\mathbf{M}_2}_{\mathbf{M}_1} = \frac{p( \mathbf{r} \ | \ \mathbf{M}_2)}{p( \mathbf{r} \ | \ \mathbf{M}_1)}.
\end{equation}
A Bayes factor $\mathcal{B}^{\mathbf{M}2}_{\mathbf{M}_1} > 1$ indicates that model $\mathbf{M}_2$ is statistically preferred over model $\mathbf{M}_1$ for the observed data. In practice, threshold values are often adopted to assess the strength of this preference. For each red-noise component, the number of Fourier modes in the spectral representation can be treated as a free parameter to better distinguish between different red-noise processes \citep{vv15}. This can either be implemented through a brute-force search for the optimal number of Fourier modes within a Bayesian model-selection framework \citep{cbp+22}, or by treating the number of Fourier modes as a hyperparameter to be inferred directly during noise modelling \citep{EPTA+2023b}.

Noise modelling relies on stochastic sampling methods to explore the parameter space. Two classes of samplers are commonly used for PTA noise analyses: Markov Chain Monte Carlo (MCMC) samplers and nested samplers. Examples of such tools include (but are not limited to) \textsc{PTMCMCSampler} \citep{ev17}, \textsc{MultiNest} \citep{fh08}, \textsc{PolyChord} \citep{hhl15}, and \textsc{dynesty} \citep{spe20}. Sampling is typically performed using uninformative priors, chosen to be uniform in either linear or logarithmic space for the noise parameters. Meanwhile, hierarchical priors based on hierarchical Bayesian modeling have been proposed recently to mitigate potential biases in noise-parameter estimation \citep{van24,gs25}. Over the past decade, most noise-modeling software packages have relied on CPU-based parallel computing (e.g., \textsc{TempoNest}, \textsc{enterprise}). GPU-accelerated toolkits are now under development to further improve the computational efficiency of PTA analyses\footnote{See e.g., \url{https://github.com/nanograv/discovery}.}.

\section{The PTA detector} \label{sec:PTA}

With the observed TOAs, timing and noise model of a number of pulsars in hand, we can now construct the full PTA detector model by adding the GW signals that we want to search for. In an ideal case if all effects on the TOAs are perfectly known, one can construct a model, including timing, noise and GW signal parameters that exactly predicts all TOAs, thus the residuals between the model and observed TOAs become zero. The detection of a GW signal then becomes a measurement on how significant it is in order for the full model to describe the observations.

\subsection{Basic properties of the detector}

\begin{figure}[!h]
    \centering
    \includegraphics[width=0.97\linewidth]{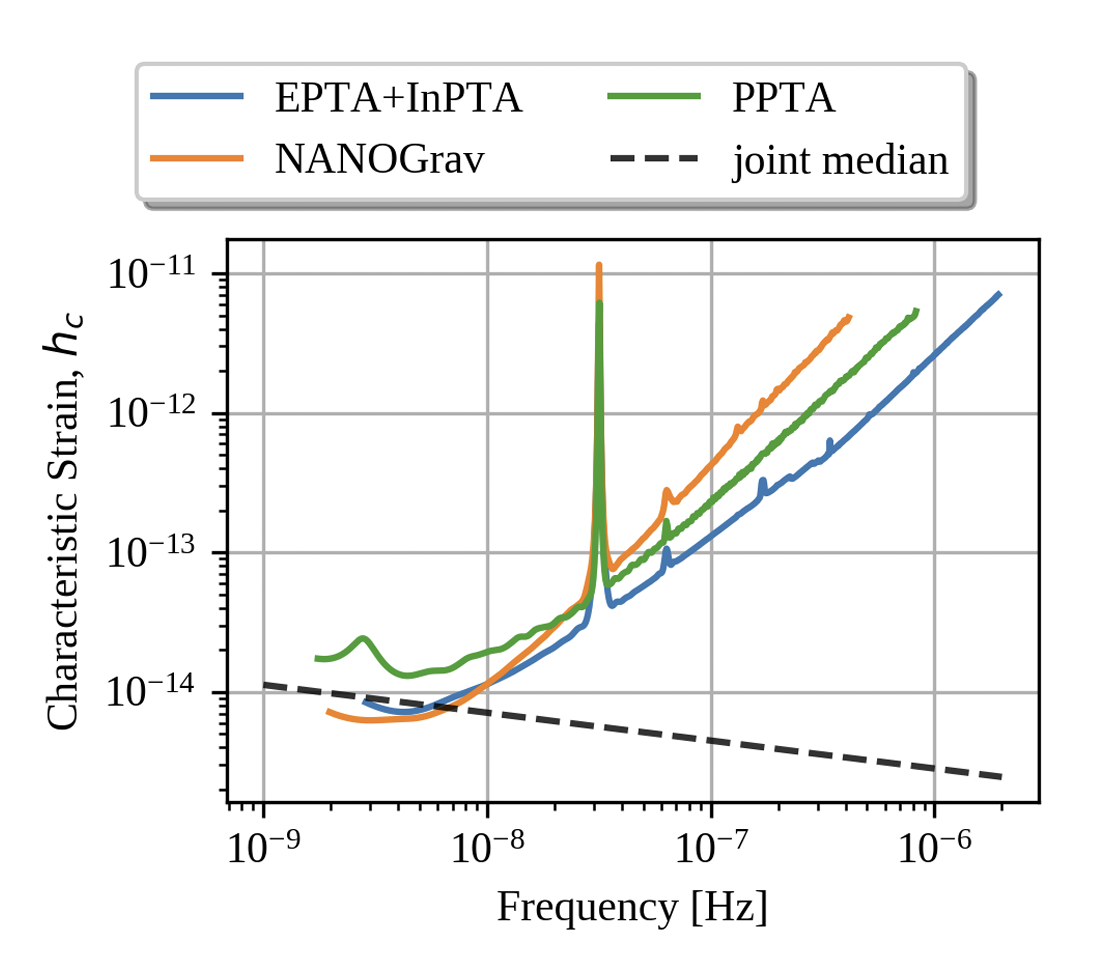}
    \caption{Sensitivity curves for the EPTA DR2, PPTA DR3 and NANOGrav 15yr data sets from the IPTA comparison \cite{2024ApJ...966..105A}}
    \label{fig:pta_sensitivity}
\end{figure}

Using some basic properties a theoretical sensitivity can be computed for a given PTA detector. These include the full observation time span $T$, observation cadence $\xi$, overall white noise level $\sigma$ and number of pulsars $N$.
Following \cite{2013CQGra..30v4015S} we can give a simple scaling relation for the signal-to-noise ratio $\rho$ of a GWB,
\begin{equation}
    \rho \approx N \frac{A^2 \xi T^\gamma}{2\sigma^2\sqrt{\gamma-1/2}} \,,
\end{equation}
where $A$ and $\gamma$ and the amplitude and spectral index of a power law describing the PSD of the GWB signal. Note that this is valid in the so-called 'weak signal' regime. If the GW signal is much stronger than the white noise, the scaling is independent of the white noise level and can be simplified to
\begin{equation}
    \rho \approx N \sqrt{\xi T} \,.
\end{equation}
In the above, we have used the simplification of knowing the pulsar noise terms exactly. A more realistic PTA sensitivity curve can be computed following \cite{2019PhRvD.100j4028H} using the best fit values for the pulsar noise models. Figure~\ref{fig:pta_sensitivity} shows the analytic sensitivity curves for the EPTA DR2, PPTA DR3 and NANOGrav 15yr data sets.

\subsection{Complete signal model of the detector}

The fundamental principle of the PTA detector is that once all known effects are taken into account, the remaining residuals should be distributed as a Gaussian centered around zero with a certain width. Including the GW signal in the model the residuals $\mathbf{r}$ can be written as
\begin{equation}
    \mathbf{r} = \mathbf{t}_{\rm obs} - \mathbf{s}_{\rm timing} - \mathbf{s}_{\rm noise} - \mathbf{s}_{\rm GW} \,,
\end{equation}
where $\mathbf{t}_{\rm obs}$ is a vector of the observed TOAs and $\mathbf{s}_{\rm X}$ is the predicted delay from the timing, noise and GW signal model respectively. A Gaussian likelihood function of the residuals can be written as
\begin{equation}
    L \propto {\rm e}^{-\frac{1}{2} \mathbf{r}^{\rm T} C^{-1} \mathbf{r}} \Big/ \sqrt{|C|}  \,,
\end{equation}
where $C$ is the full covariance matrix. It consists of both the noise terms and the GW signal including spatial correlations. The timing model design matrix can be marginalized and thus does not appear in this covariance matrix. A more detailed review on the likelihood function can be found in \cite{epta6_gwb}.

Using this likelihood function, one can perform both frequentist and Bayesian analyses to determine the evidence for a model and perform model comparisons. To estimate the significance of a model with a GW signal its evidence needs to be compared against a distribution of evidences from data sets that do not have the GW signal, i.e., a null-distribution. Only if the significance is large enough can a claim of detection be made.

The PTA community has managed to gather various analysis scripts used for the earliest results and to create a framework to efficiently compute the likelihood function. The framework is designed to be modular and able to include new terms of noise and signals \citep{evt+19}. \textsc{enterprise} has since then become the standard analysis software for GW searches with PTAs. As the PTA data volume is slowly increasing with more observations and more pulsars added, the need for an even more efficient software package has risen. \textsc{discovery}\footnote{\url{https://github.com/nanograv/discovery}}, which uses GPU-acceleration, is being developed and could take over from \textsc{enterprise} as the next generation standard PTA analysis software package.

\subsection{Detecting a gravitational wave background}

As a GWB affects all pulsars, a conclusive detection will include two parts, the common red signal (CRS) and the characteristic spatial correlation. The CRS is the more prominent of the two and should be measured first. A simple description of the GWB is given by a power law model,
\begin{equation}
    S_{\rm GWB}(f) = \frac{A_{\rm GWB}^2}{12\pi^2} \left(\frac{f}{f_{\rm r}} \right)^{-\gamma_{\rm GWB}} {\rm yr}^3 \,,
\end{equation}
where $A_{\rm GWB}$ is the amplitude and $\gamma_{\rm GWB}$ is the spectral index. The power law amplitude $A_{\rm GWB}$ is related to the GW characteristic strain $h_c$ and energy density spectrum $\Omega_{\rm GWB}$ as follows:
\begin{align}
    h_c(f) & = A_{\rm GWB} \left( \frac{f}{f_{\rm r}} \right)^{\alpha_{\rm GWB}} \,, \\
    \Omega_{\rm GWB}(f) & = \frac{2\pi^2}{3H_0^2} f^2 A_{\rm GWB}^2 \left( \frac{f}{f_{\rm r}} \right)^{2\alpha_{\rm GWB}} \,,
\end{align}
where $\alpha_{\rm GWB}=\frac{3-\gamma_{\rm GWB}}{2}$ is a different spectral index and $H_0$ is the Hubble constant.
However, several sources of noise can create a CRS, e.g. clock errors, SSE systematics or other sources of pulsar timing noise (see Section~\ref{ssec:noise}). Therefore, the spatial correlations provide conclusive evidence of the origin of the CRS. For an isotropic GWB with GR as the theory of gravity, the characteristic spatial correlations only depend on the angular separation of the pulsars on the sky, it is also known as the HD correlation \cite{hd83}.
\begin{equation}
    \mathbf{\Gamma}(\zeta_{ij}) = \frac{3}{4} (1-\cos \zeta_{ij}) \ln \left( \frac{1-\cos \zeta_{ij}}{2} \right) - \frac{1-\cos \zeta_{ij}}{8} + \frac{1}{2} + \frac{1}{2}\delta_{ij} \,,
\end{equation}
where $\zeta_{ij}$ is the angular separation on the sky between two pulsars $i$ and $j$ and $\delta_{ij}$ is the Kronecker delta symbol, i.e., 1 if $i=j$ and 0 otherwise. This overlap reduction function (ORF) is characteristic for a GWB and depends only on the angles of the pulsar pairs.

Although a large amount of evidence can be found, the significance of the evidence can only be estimated through comparison to a null-distribution. For a GWB, the signal model includes the CRS as well as the HD correlations. A conservative null-hypothesis model should include the CRS, but not the characteristic spatial correlations. This type of CRS could arise from pulsar noise or other sources.

As our Universe is one realization of many possible universes, the observed PTA data set will either be affected by a GWB or not. Assuming that it does, to construct data sets without a GWB we need to either modify the real PTA data set or simulate realistic data sets without a GWB. Two methods have been proposed to remove the HD correlation from a PTA data set while preserving the pulsar noise: sky scramble and phase shifts. However, in practice, these two methods may not be able to fully remove all correlations \citep{2023ApJ...956...14D}. Therefore, simulated data sets may be better in constructing null-distributions \citep[e.g.,][]{2023PhRvD.108j4050H,2025PhRvD.112j3009V}. The estimated significance of a GWB signal could, thus, depend on the method used to construct the null-distribution. Future investigations will allow us to be more confident in the significance of a GWB signal.

\subsection{Detecting continuous gravitational waves}

If an individual SMBHB is loud and close enough to Earth, the GWs emitted could be detected by PTAs. Since the observation duration of PTAs of decades is vastly shorter than the merger time of SMBHBs of millions of years, they can be assumed to emit continuous gravitational waves (CGWs). The effect of the GWs on the TOAs can be computed deterministically for all pulsars in the array assuming that the basic properties of the source and its location are fully known. The delays caused by the SMBHB can be split into two parts, one at the Earth and the other at the pulsar:
\begin{equation}
    s(t,\hat{n}) = \sum_{A=+/\times} F^A(\hat{n}) [s_A (t) - s_A (t-\tau)]
\end{equation}
where $\hat{n}$ is the direction of the propagation of the GWs, $A=+/\times$ refers to the two polarizations of the GWs, $F^A$ are the antenna pattern functions and $s_A(t)$ and $s_A(t-\tau)$ are the Earth and pulsar term respectively. For the full expressions of $F^A$ and $s_A$ and more details on CGWs, see e.g., \cite{2023MNRAS.521.5077F}. 

To calculate the delays, the exact distances between the source and the pulsars are also required. However, as we do not have accurate measurements of them, they are often considered as a possible source of noise. Since our measurements will improve over time, so can we reduce the noise.

The second implicit assumption is that the SMBHB evolves very slowly compared to the observation length of a few decades. This translates into assuming that the frequency at which the GW is emitted does not change much during the PTA observation, thus it is monochromatic and can be approximated as a sine wave.

Finally, we assume that the SMBHB evolves purely from GW emission and has reached a circular orbit when it enters the PTA frequency range. If the binary is eccentric or other mechanisms still play an important role in its evolution, the energy emitted in GWs could be split into many different frequencies. Then, the sine wave approximation will be invalid and the waveform, i.e., the induced delays will become more complex. Astrophysically, there are possible mechanisms that drive the evolution of the SMBHB, which could leave a remaining eccentricity or environmental impact on the binary. Thus, the PTA community is also searching for hints of these effects.

\begin{figure}[!h]
    \centering
    \includegraphics[width=0.97\linewidth]{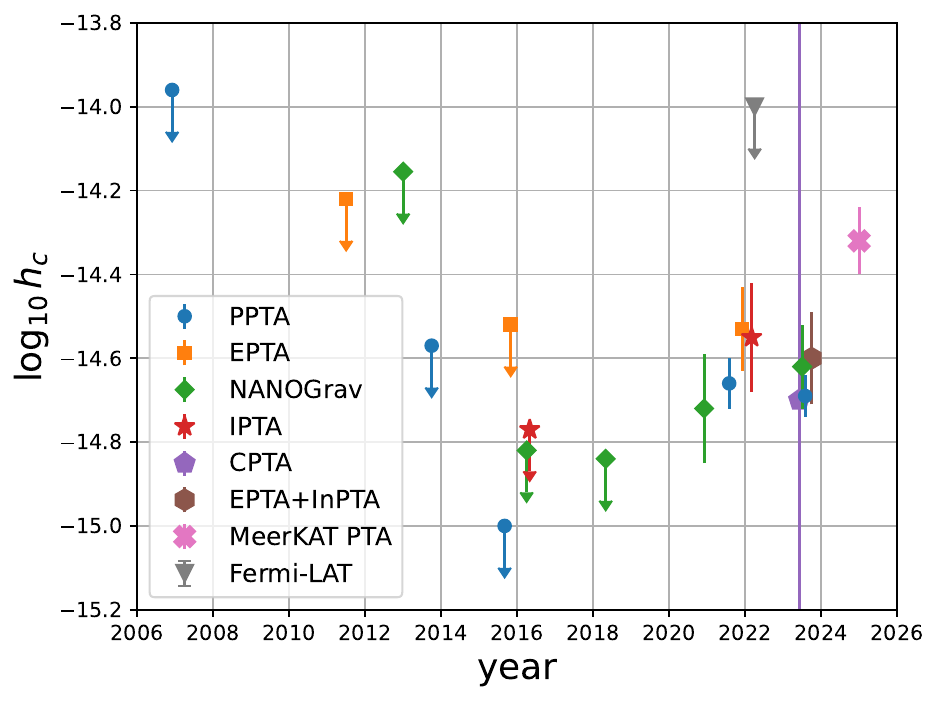}
    \caption{Evolution of the constraints, upper limits and measurements, on the GWB amplitude at $\gamma_{\rm GWB}=13/3$ over the last two deacdes, figure adapted from \cite{2024ResPh..6107719V}. The upper limit from the Gamma-ray PTA using Fermi-LAT data \cite{2022Sci...376..521F} and the constraint from the MeerKAT PTA \cite{meerkat_gwb} have also been added.}
    \label{fig:pta_review}
\end{figure}

\section{Evolution of PTA results in the last decade} \label{sec:result}

In the last decade, the PTA community has made tremendous achievements towards the detection of nanohertz GWs. Starting from upper limits to the detection of a common red signal and finding evidence that the signal could indeed be from GWs.

Figure~\ref{fig:pta_review} summarizes the evolution the GWB measurements over the last years \cite{2024ResPh..6107719V}. The early era of PTAs from 2006-2020 can be seen as the upper limit era, where the addition of more and better data helped to improve the overall sensitivity of the PTA. This allowed us to set increasingly stringent upper limits. 
The first claim of measuring a common red signal in 2020 broke this trend, starting the era of common red signals \citep{abb+20,ccg+21,gsr+21}. This is the first major achievement of the community in the quest to detect nanohertz GWs. With a signal in hand, interpretation efforts could start in earnest. 2023 marked the achievement of the second milestone with the measurement of significant evidence for the common red signal to be of GW origin, while the nominal amplitude remained consistent with the previous measurements. This evidence of a GW signal at nanohertz has also sparked interest from other communities, in particular exploring the possible source of such a signal. As we move forward, we hope to enter the next era of a confirmed GW detection, and work on the characterization of the possible source or sources.

\subsection{Progress in precision timing and noise characterization}

In order to robustly detect GWs, the timing and noise properties of each pulsar in the array have to be modelled very precisely. While the deterministic timing model is usually well understood, the stochastic noise in the timing data has so far mostly been modelled phenomenologically. Especially, the covariance between any GW signal and other noise components is usually significant. As pulsar timing has become increasingly precise over the last decade, the focus has shifted towards more robust and physically motivated noise models \cite{ng156_noise,ng12.5_noise}.

The standard PSD for time-correlated noise is the power law model, which has been shown to fit the PTA data reasonably well enough for GW searches. However, in recent years, several optimization methods have been developed and applied to the PTA analysis to diminish biases and other issues that could come from using the power law model. As discussed in Section~\ref{ssec:noise}, these include determining the optimal choice of frequency binning \citep{epta6_noise,eptadr2_noise,cptadr1_noise}, adding more physically motivated noise processes such as scattering variation \citep{eptadr2_noise,inptadr1_noise,inptadr2_noise}, solar wind \citep{2022ApJ...929...39H,2025A&A...704A.109I} and time-domain events \citep[e.g.,][]{ssj+21,jcc+24}, accounting for long-term instability of some legacy observing systems or at certain observing frequency bands \citep{iptadr1_noise,pptadr2_noise,pptadr3_noise} and so forth.

\begin{figure}[h!]
    \centering
    \includegraphics[width=0.975\linewidth]{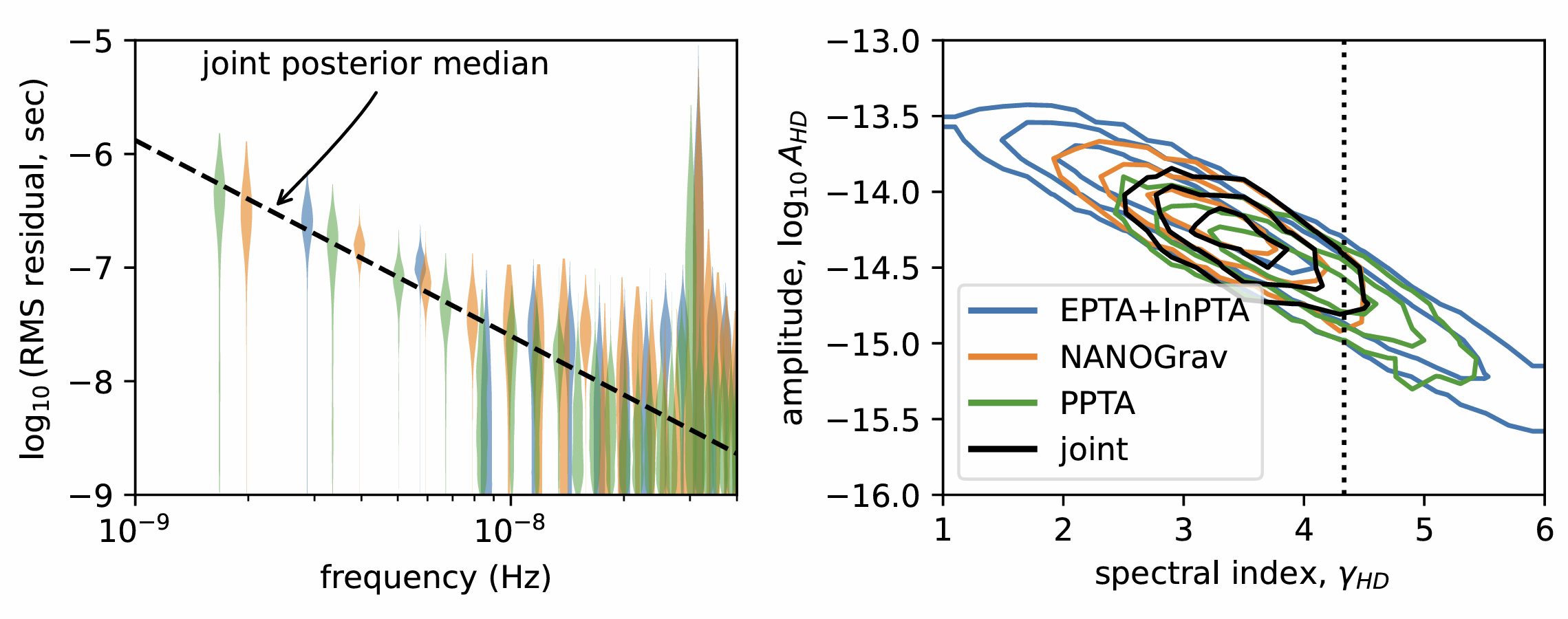}
    \caption{GWB constraints from the EPTA DR2, PPTA DR3 and NANOGrav 15yr data sets from the IPTA comparison \cite{2024ApJ...966..105A}.}
    \label{fig:ipta_comp}
\end{figure}

\begin{figure}[h!]
    \centering
    \includegraphics[width=0.45\linewidth]{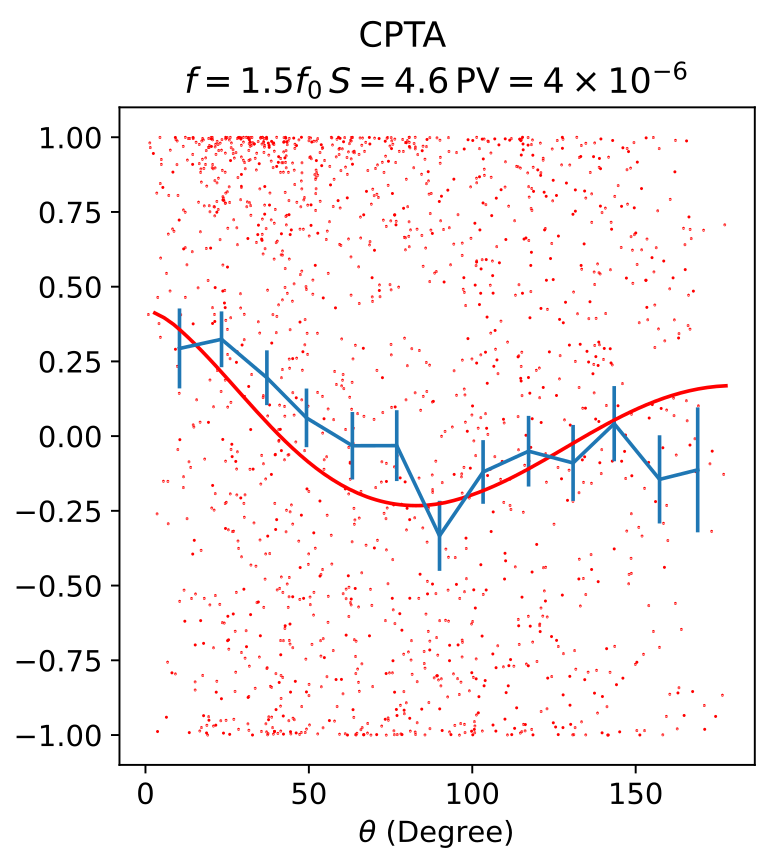}\\
    \includegraphics[width=0.45\linewidth]{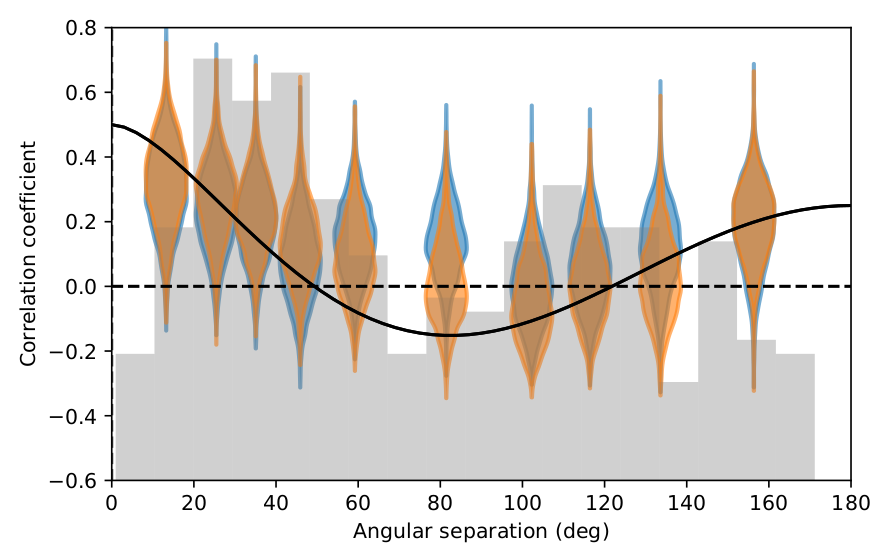}
    \includegraphics[width=0.45\linewidth]{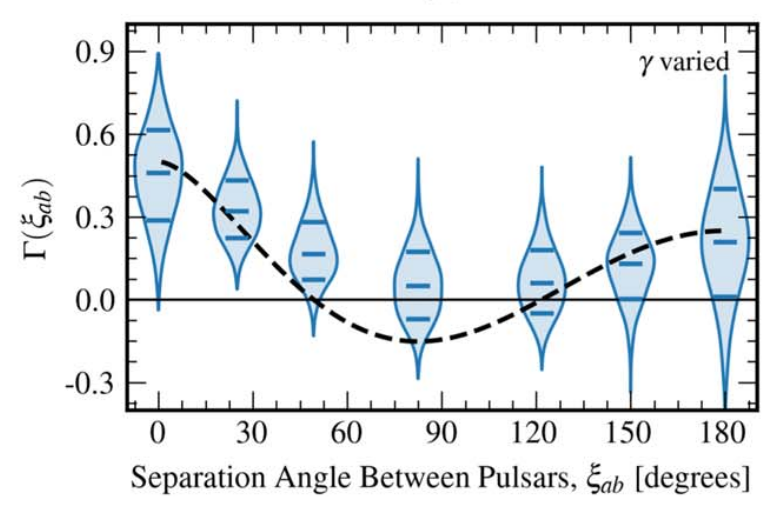}
    \includegraphics[width=0.45\linewidth]{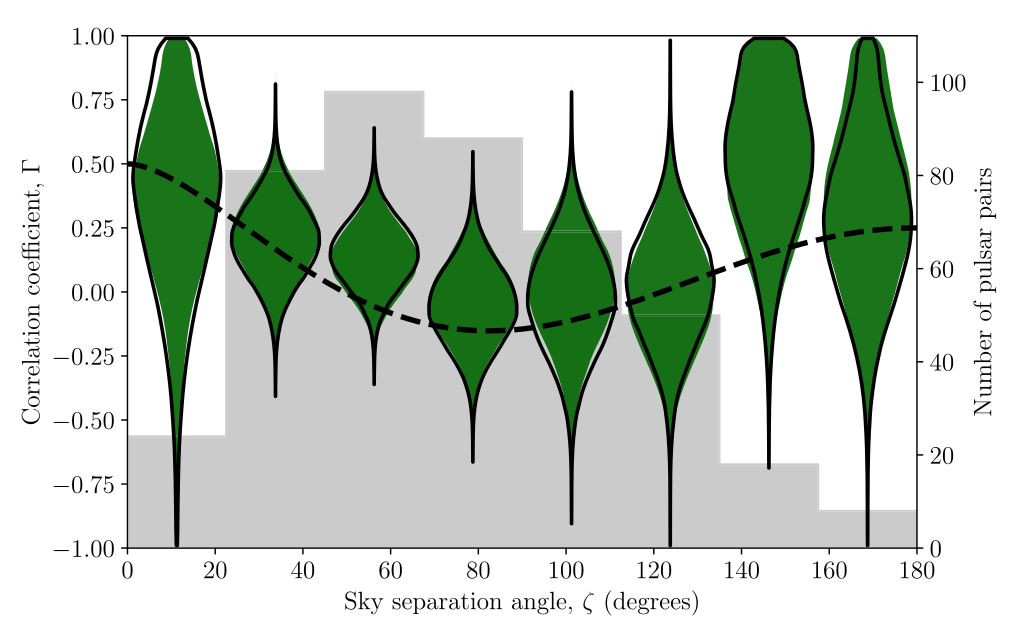}
    \includegraphics[width=0.45\linewidth]{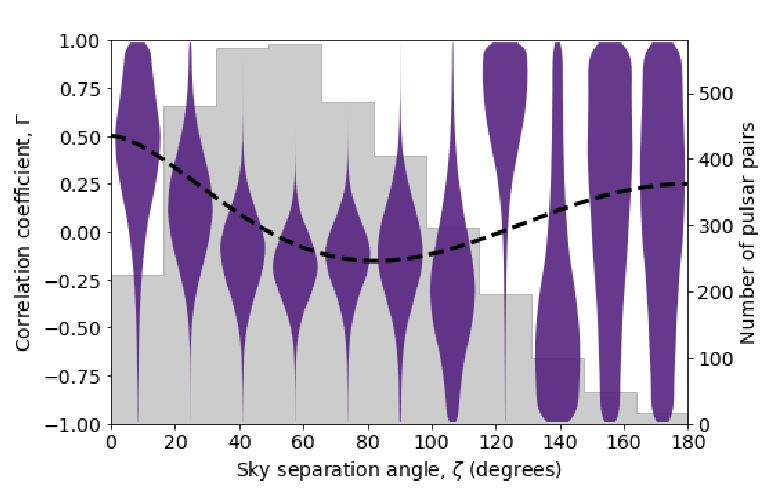}
    \caption{The recovered overlap reduction function from the CPTA DR1, EPTA DR2, NANOGrav 15-year, PPTA DR3 and MeerKAT PTA data sets (from top left to bottom right) compared against the Hellings and Downs curve.}
    \label{fig:orf}
\end{figure}

\subsection{Gravitational wave background}

The GWB is one of the sources that PTAs can probe. A background can originate either from the incoherent sum of a very large number of individual sources or from a cosmological source. The advantage of this source in the PTA range is that the significance of the signal will slowly build up over time as more observations are taken. In particular, this can clearly be seen from the progress the PTA community has achieved over the last decade.

\subsubsection{Upper limits}

In the initial phase when PTAs started operating as experiments the quality of the MSP observations lacked the sensitivity to measure a weak GW signal. However, as time went on, both the increase in number of observations and pulsars as well as instrumental advancements have quickly improved the sensitivity. Nonetheless, even without measuring a signal, one can put upper limits on the strength of the signal. As the first data sets were released and analyzed, the upper limit quickly decreased to a level that is meaningful for astrophysical interpretation efforts.

In 2015--2016, the upper limits on GWB from the EPTA \citep{2015MNRAS.453.2576L}, PPTA \citep{2015Sci...349.1522S}, NANOGrav \citep{2016ApJ...821...13A} and the IPTA \citep{2016MNRAS.458.1267V} were within the range of $1--3\times10^{-15}$, depending the number of pulsars and modeling complexity employed in the analysis.
With the EPTA having seen a small hint of a signal pushing the upper limit to a larger value, the collaboration has focused on improving the instruments and gaining a deeper understanding of their data set. NANOGrav has been working on adding more data to increase the sensitivity as well as improving the data analysis methods. The 11-year NANOGrav data set was the first to use a SSE model to set the upper limit \citep[$1.45\times10^{-15}$,][]{2018ApJ...859...47A}. However, as the data set increased in sensitivity, eventually the upper limit began to stagnate and in 2020 with the 12.5-year data set NANOGrav found conclusive evidence for a common red signal \citep{2020ApJ...905L..34A}.

\subsubsection{Detection of a common red signal}

After the detection of a CRS with the NANOGrav 12.5-year data set \citep{2020ApJ...905L..34A}, the PPTA and EPTA collaborations were able to confirm the existence of this CRS with the PPTA DR2 \citep{pptadr2_gwb} and EPTA DR2 6 pulsar data sets \citep{epta6_gwb}. The PTA community came together and finished the analysis of the IPTA DR2 concluding with another confirmation of the CRS at a consistent level \citep{iptadr2_gwb}. This has marked the year 2020 and 2021 as two years in which PTAs have found a signal with an amplitude and spectrum that could be of GW origin and therefore could be used for interpretation work.

The nominal level of the CRS was found to be around $2.0\times10^{-15}$. Although the nominal amplitude is an initial indicator of the strength and origin of the GWB signal, the spectrum of the signal across a frequency range contains more useful information to identify and characterize its source. Thus, it is more relevant to compare the amplitude and spectral index of the power law models or the frequency resolved power of the GWB.

Nonetheless, the CRS is only the first part of the full GWB signal, as it only uses the auto-terms of the signal between pulsars. Such a common signal can also arise from noise sources other than GWs, e.g. clock errors, SSE systematics or common pulsar noise properties. Thus, in order to be able to claim a robust detection of GWs, we need to find the second part of the GWB signal, i.e., the characteristic spatial correlations, the HD curve.

\subsubsection{Evidence of gravitational wave origin}

The latest milestone was achieved in 2023 when using the most recent data sets the EPTA+InPTA \citep{eptadr2_gwb}, NANOGrav \citep{ng15_gwb}, PPTA \citep{pptadr3_gwb} and CPTA \citep{cpta_dr1} have all found evidence for the HD correlation, ranging from $\sim 2 - 4.6 \sigma$. Thus, the CRS is indeed very likely of GW origin. This is yet another major step forward for the community and also shows that the last decade has been very fruitful for the search of nanohertz GWs with PTAs. In 2024, the MeerKAT PTA collaboration provided yet another independent confirmation of significant evidence of $\sim 2 - 4 \sigma$ for a GW signal depending on the pulsar noise model \cite{meerkat_gwb}.

Figure~\ref{fig:ipta_comp} shows a comparison of the GWB constraints from the 2023 series of results, while Figure~\ref{fig:orf} shows the measured ORF $\mathbf{\Gamma}(\zeta_{ij})$ of the various PTA data sets between the pulsar pairs compared to the HD correlation curve. The consistency between the different posterior distributions as well as the matching of the measured ORFs with the HD curve indicate a consistent measurement of a GWB signal across different independent data sets and analyses \citep{2024ApJ...966..105A}.

Currently, the PTA community is working towards assembling and analyzing the next IPTA data combination, the third data release (DR3), which will combine data sets from all regional PTA collaborations to create the most sensitive PTA data set to date. With over 100 pulsars and two decades of observations, we hope to be able to provide conclusive evidence for the detection of nanohertz GWs.

\begin{figure}
    \centering
    \includegraphics[width=0.8\linewidth]{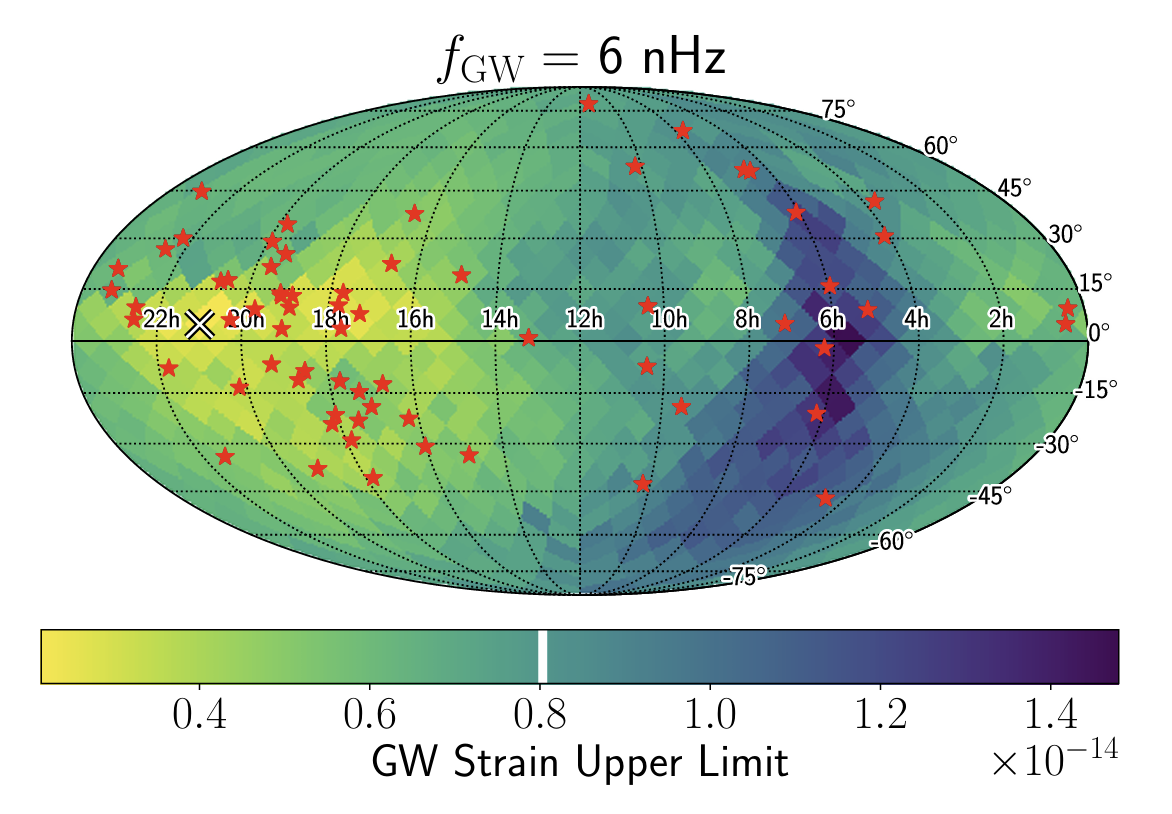}
    \includegraphics[width=0.75\linewidth]{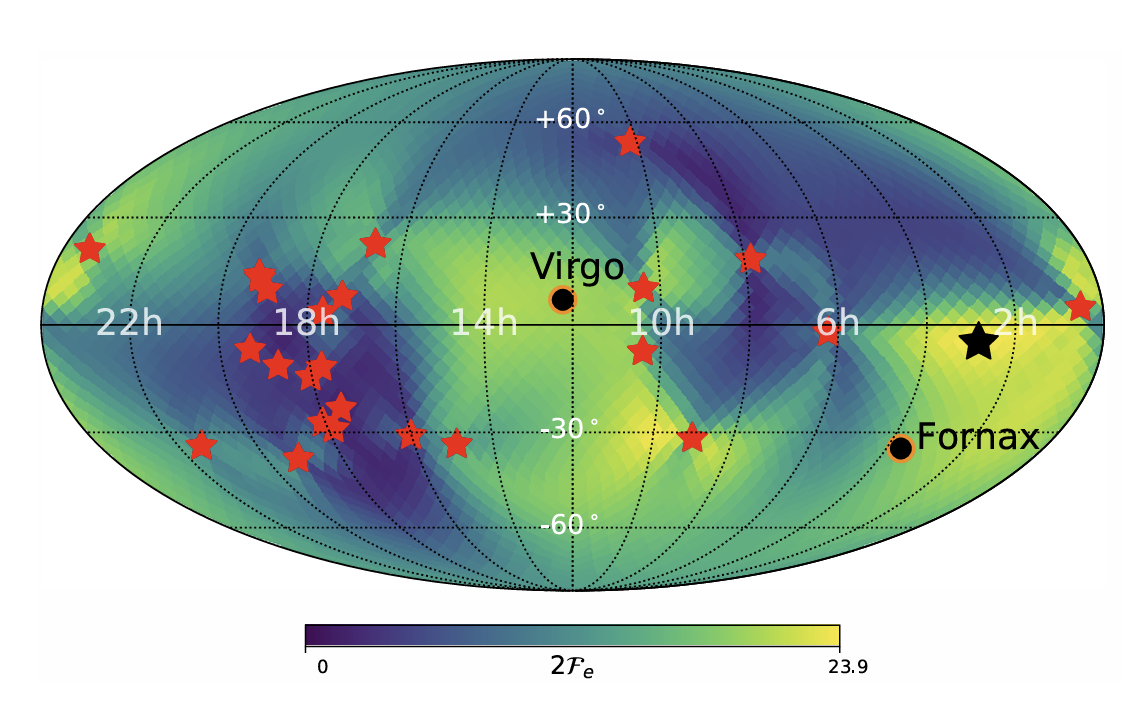}
    \includegraphics[width=0.8\linewidth]{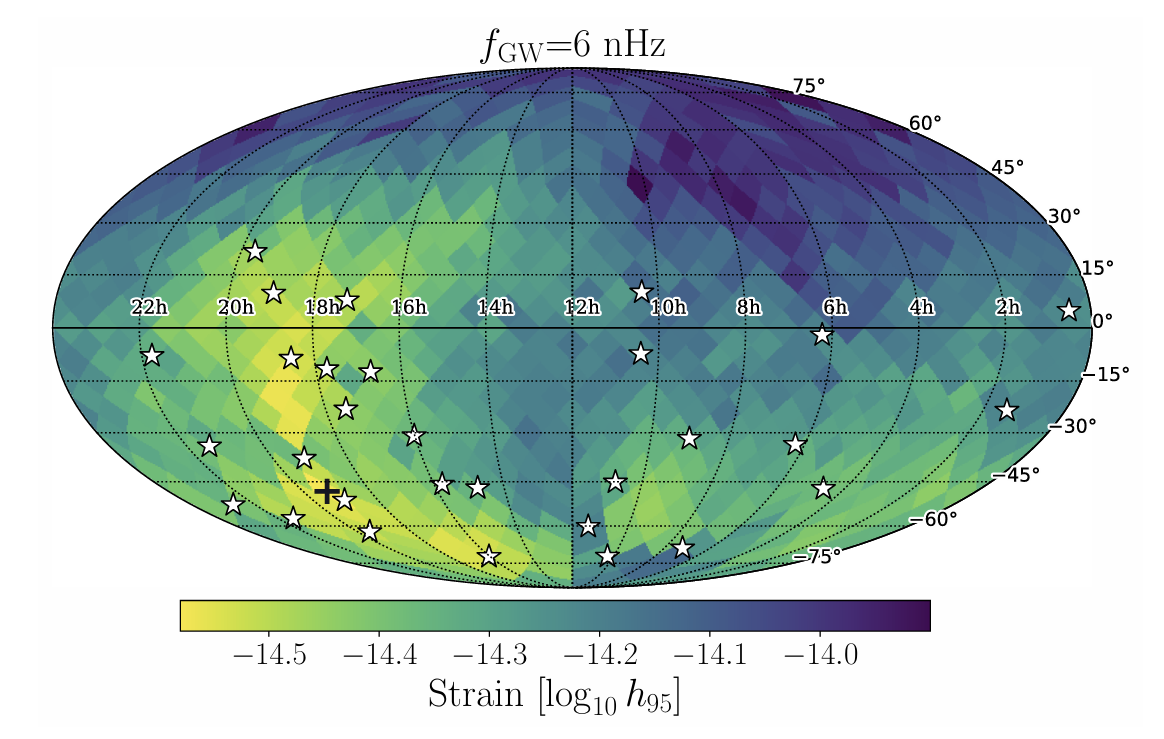}
    \caption{Sky maps of the constraints from the search for individual SMBHBs with the NANOGrav 15-year (top) \cite{2023ApJ...951L..50A}, EPTA DR2 (middle) \cite{2024A&A...690A.118E} and PPTA DR3 (bottom) \cite{2025ApJ...992..181Z} data sets.}
    \label{fig:cgw}
\end{figure}

\subsection{Continuous gravitational waves}

Aside from the GWB, PTAs have also been searching for individual sources of GWs. The most prominent source of which are individual super massive black hole binaries (SMBHBs) from galaxy mergers. In contrast to a GWB the distribution of CGWs is more dependent on the realization of our Universe. Thus, one or many strong CGW sources could fortunately be detectable by PTAs. The searches in the last decade have set more and more stringent upper limits on SMBHBs in our vicinity. With the finding of evidence of a signal of GW origin, we have reached a stage of stagnant upper limits and possible hints for individual CGW sources.

\subsubsection{From upper limits to hints}

The first search for individual SMBHBs were performed in 2014 by NANOGrav \cite{2014ApJ...794..141A} and the PPTA \cite{2014MNRAS.444.3709Z}, setting the first upper limit on CGWs, since no evidence was found. This work was followed by the analysis from the EPTA using the DR1 in 2016 \cite{2016MNRAS.455.1665B}. These upper limits pushed the constraints from about $3.0\times10^{-14}$ to $1.7\times10^{-14}$ and $1.5\times10^{-14}$ at the most sensitive frequency. The sky maps are dominated by the distribution of pulsars on the sky, i.e., it is more sensitive where there are more pulsars.

With the measurement of a CRS in the 12.5-year data set, the searches for individual CGW sources have become more complex. Adding a CRS as a source of noise into the search for individual SMBHBs with the 12.5-year data set NANOGrav set an upper limit of $6.82 \times 10^{-15}$ \cite{2023ApJ...951L..28A}, which is similar to the results from the 11-year data set \cite{2019ApJ...880..116A}. As the CRS has become a standard addition to the model, the IPTA DR2 has also been searched for individual CGW sources. With no evidence found, an upper limit was set at $9.1\times10^{-15}$ \citep{2023MNRAS.521.5077F}.

The 2023 release of coordinated papers includes searches for individual SMBHBs from both the EPTA \citep{2024A&A...690A.118E} and NANOGrav \citep{2023ApJ...951L..50A}. Both collaborations have found some consistent hints of a CGW source at $\sim$4 nHz. This possible source is very poorly localized and lacks strong enough evidence to claim a detection. In contrast, the analysis of the PPTA DR3 does not show a strong significance for a CGW source at any frequency resulting in an upper limit of $7.0\times 10^{-15}$ \citep{2025ApJ...992..181Z}. Sky maps of the upper limits for these analyses can be found in Figure~\ref{fig:cgw}.

\subsubsection{Correlation with the GWB}

While the earlier searches for CGWs were performed without adding a CRS in the analysis model, the searches after the measurement of the CRS have added this signal into the model. This is also more astrophysically motivated as the individual SMBHBs are forming the population for the GWB.

The results from the EPTA DR2 \citep{2024A&A...690A.118E} and NANOGrav 15-year CGW searches \citep{2023ApJ...951L..50A} have found marginal evidences for a CGW when using a CRS without spatial correlations. These evidences, however, vanish when using the HD correlated GWB model. This finding confirms the covariance between the GWB and an individual CGW source. With more data, better analysis methods and external observations, this degeneracy could be broken in the future.

\begin{figure}
    \centering
    \includegraphics[width=0.97\linewidth]{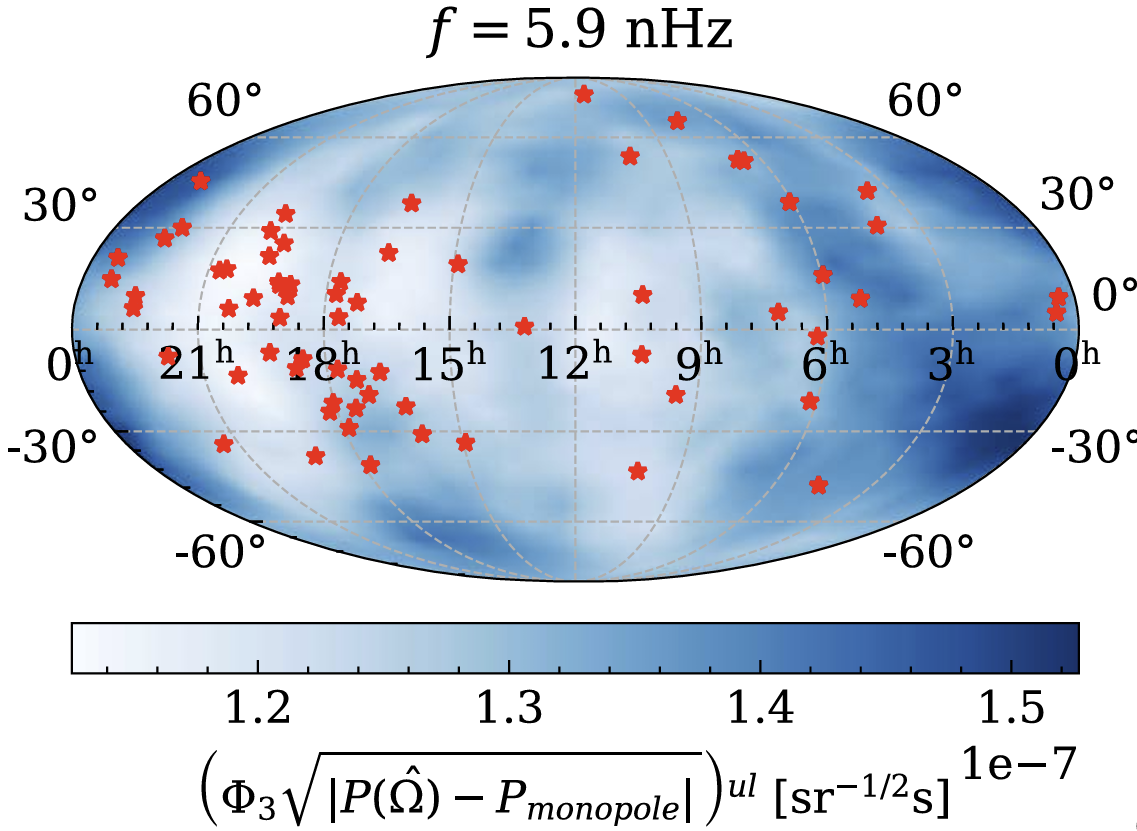}
    \includegraphics[width=0.97\linewidth]{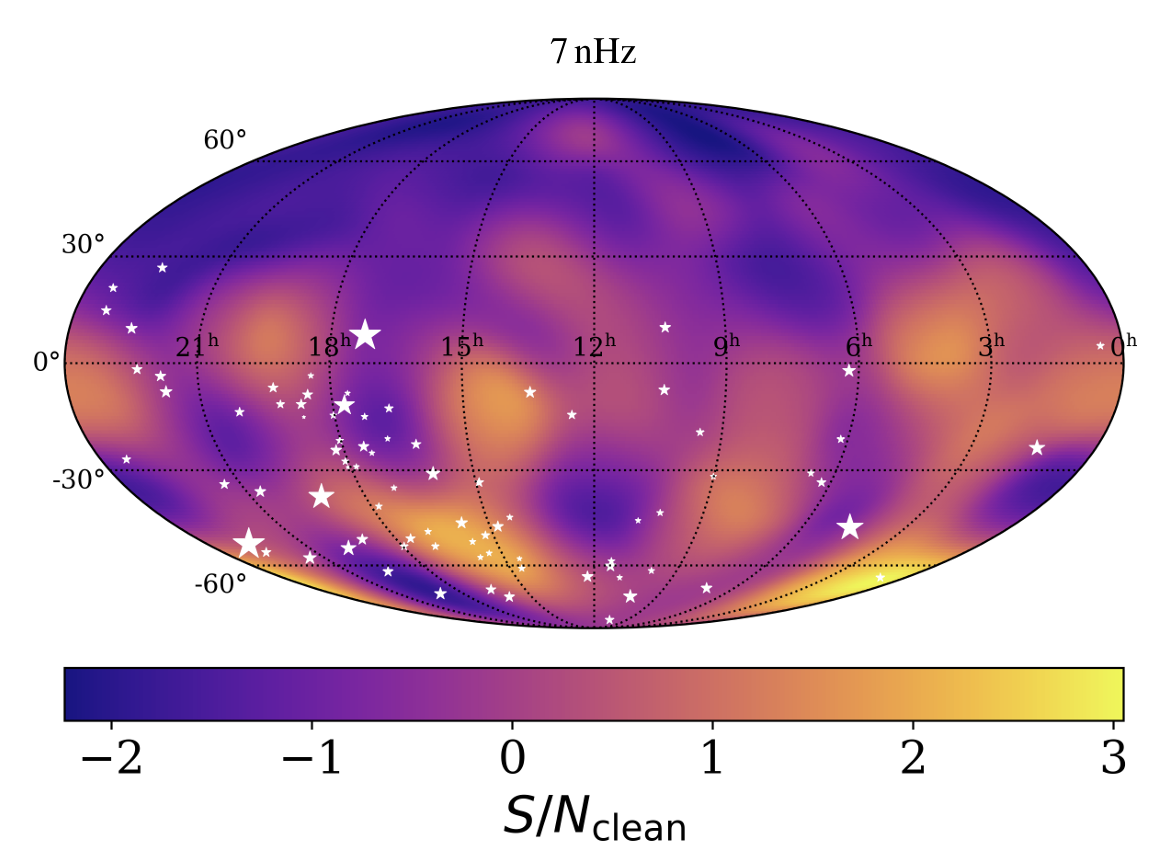}
    \caption{Sky maps of the anisotropy constraints from the search with the NANOGrav 15-year (top) \cite{2023ApJ...956L...3A} and MeerKAT PTA (bottom) \cite{2025MNRAS.536.1501G} data sets.}
    \label{fig:anisotropy}
\end{figure}

\subsection{Anisotropy of the GWB}

As the residuals measured by PTAs is the total sum of all GWs in the local Universe, background and foreground sources may all be present in the data sets. Thus, searches for anisotropies have become important for PTAs in the last decade. The level of anisotropy can provide further indication of the source of the background signal. A bright spot may hint at a resolvable single source.

Different PTA collaborations have performed anisotropy studies. The results from NANOGrav 15-year data set have lowered the strength of anisotropies in the GWB to below 20\% that of the monopole. This improves on a previous limit of about 40\% from the EPTA \cite{2015PhRvL.115d1101T}. While no evidence for a hot spot was found in the NANOGrav 15-year data set \cite{2023ApJ...956L...3A}, the MeerKAT PTA found a hot spot with a p-value of $p=0.015$ \cite{2025MNRAS.536.1501G}. Figure~\ref{fig:anisotropy} compares the sky map of the NANOGrav 15-year data set with that from the MeerKAT PTA, with the hot spot in the bottom right area of the sky map. While this is an curious finding, the significance is still very small. 

\begin{figure}[!h]
    \centering
    \includegraphics[width=0.97\linewidth]{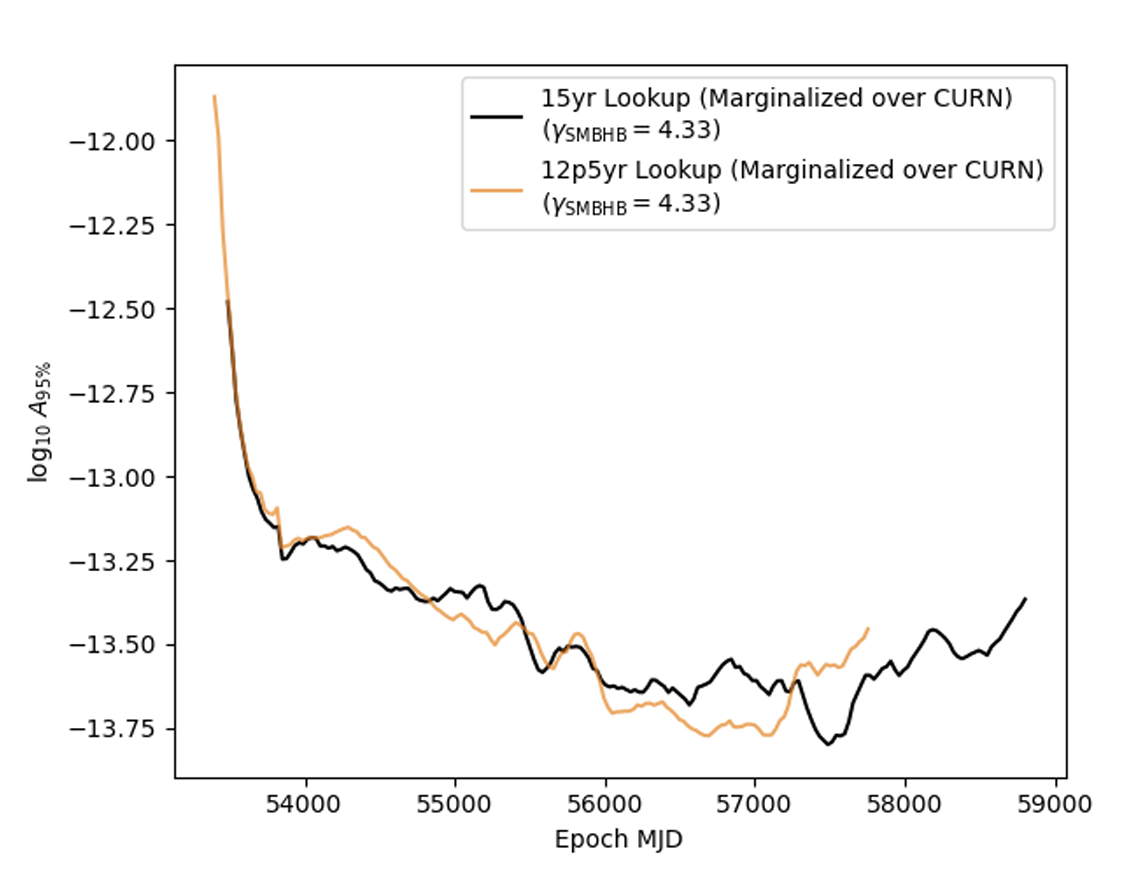}
    \includegraphics[width=0.93\linewidth]{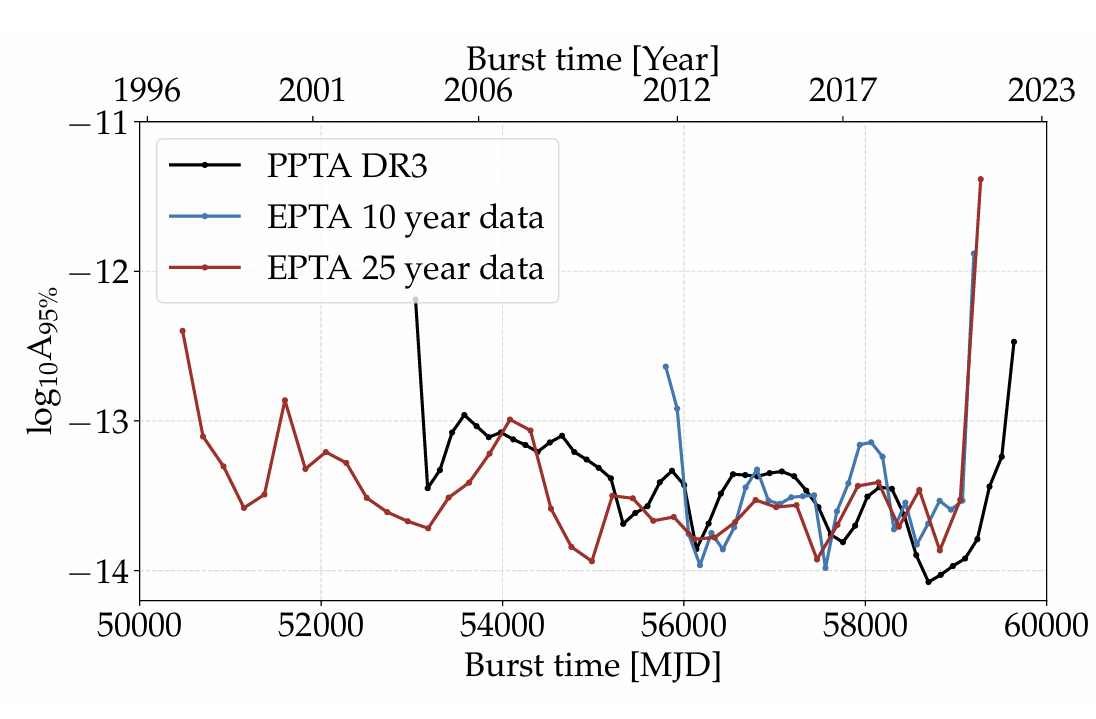}
    \caption{Upper limits on the amplitude of a GW burst with memory as a function of time from the search with the NANOGrav 12.5/15-year (left) \cite{2025ApJ...987....5A} and  EPTA/PPTA (right) \cite{2026ApJ...996L...9T} data sets.}
    \label{fig:memory}
\end{figure}

\subsection{Burst with memory}

The third and final category of sources can be classified as transient events. Although it is unlikely that we can detect any individual merger events, some events can leave signatures imprinted into spacetime that could be detected with PTAs. One such signature is a permanent deformation caused by the merger of two SMBHBs. This type of merger may cause an offset in the waveform of the signal after the merger, which corresponds to a linear trend in the PTA residuals. The unambiguous detection of such an event requires a strong signal that has a robust significance using different pulsars, data sets, models and so on.

Searches for such an event have been performed for various data sets. While no definitive signal was found, constraints have been set on the amplitude of such a jump. The first upper limits for a burst with memory were set at $2.4\sim 10^{-13}$ \cite{2015MNRAS.446.1657W} and $\sim 10^{-13}$ \cite{2015ApJ...810..150A}. Over the last decade, this has improved to about $3.3 \times 10^{-14}$ averaged across the sky \cite{2024ApJ...963...61A}. The 15-year data set \cite{2025ApJ...987....5A}, however, does not deliver much better constraints as shown in Figure~\ref{fig:memory}. This could be due to the presence of the CRS signal. A search was also carried out on EPTA DR2 and PPTA DR3 giving upper limits of order $\sim 10^{-14}$ \cite{2026ApJ...996L...9T}, which can be seen in Figure~\ref{fig:memory}. Although some hints have been seen in these analyses, they are still consistent with random noise fluctuations.

\section{Original of the GW signal} \label{sec:origin}

The GW signal from PTAs can provide valuable information into the physics of the Universe. Assuming that we have found a GW signal, we can use the properties to study the underlying source. The detection of a GW signal will open another window of the GW and multi-messenger astronomy. In this section, we present some of the most promising sources as well as some fundamental science PTAs can do with the GW signal. There are two broad categories of sources: 1. Astrophysical, likely super massive black hole binaries and 2. Cosmological, sources from the Early Universe, including exotic sources.

\begin{figure}[!h]
    \centering
    \includegraphics[width=0.52\linewidth]{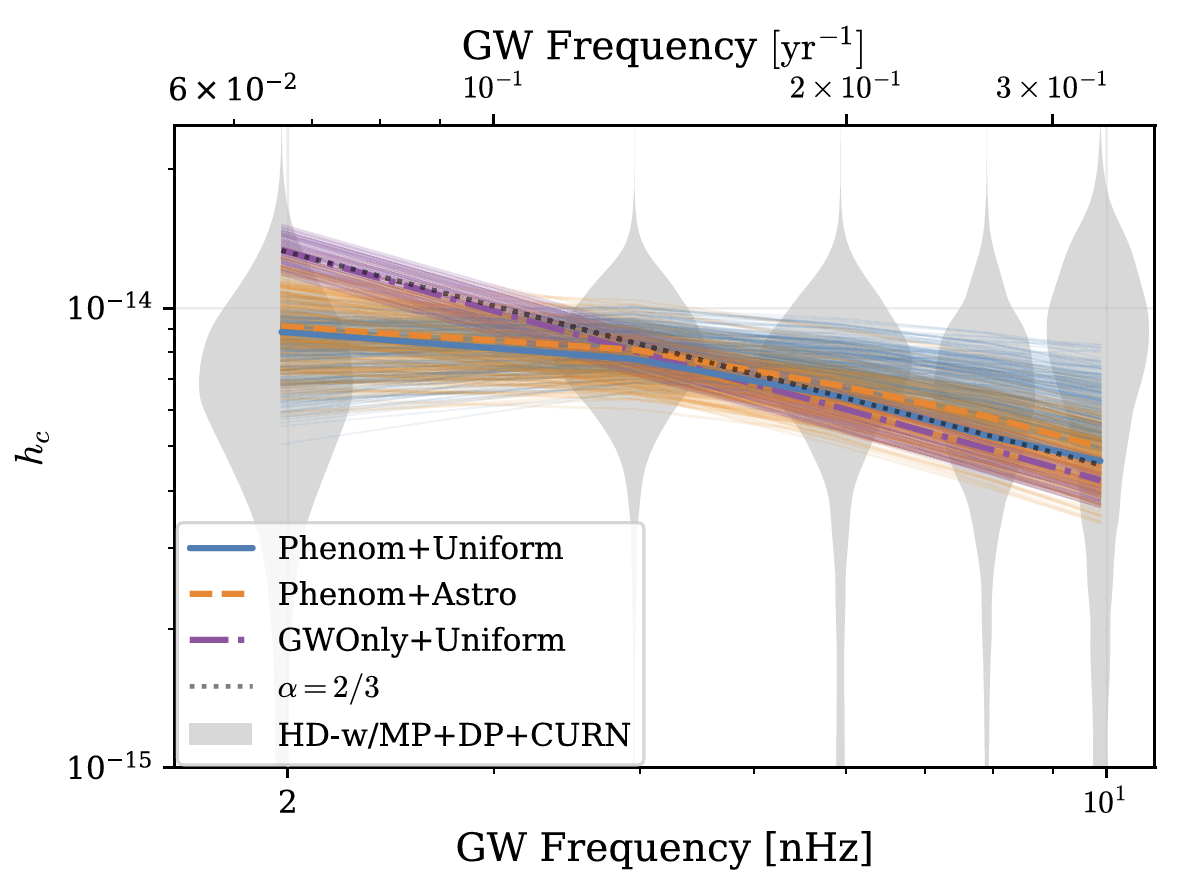}
    \includegraphics[width=0.46\linewidth]{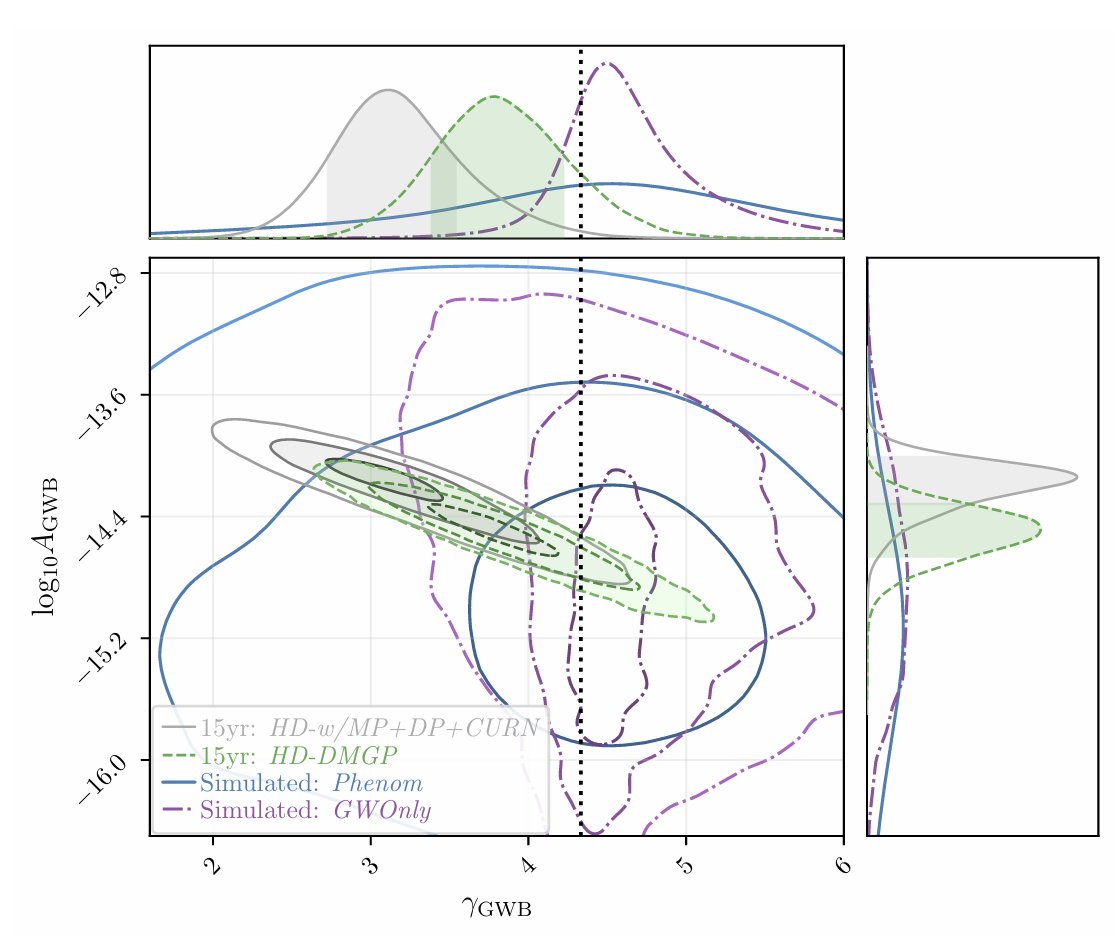}
    \includegraphics[width=0.55\linewidth]{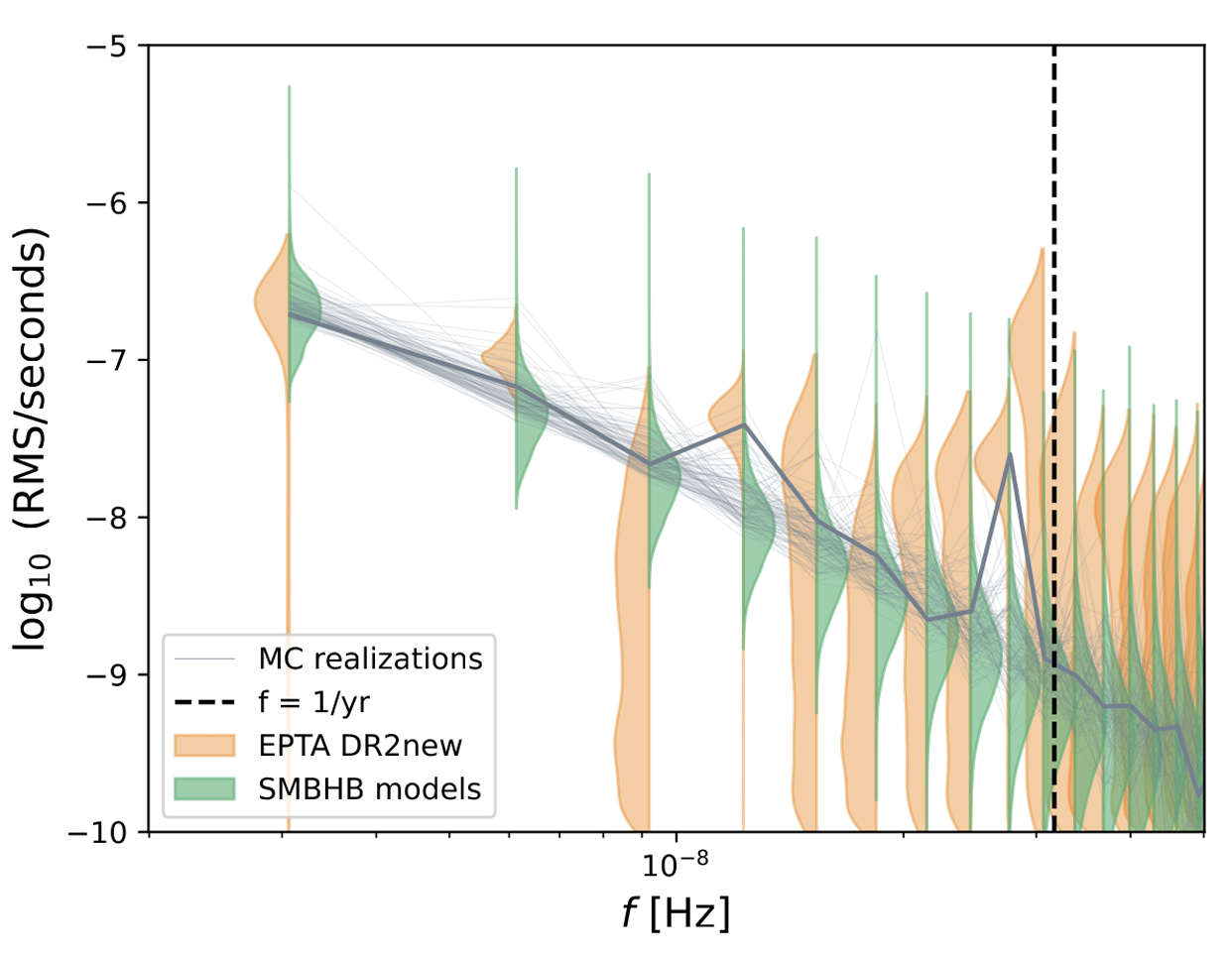}
    \includegraphics[width=0.43\linewidth]{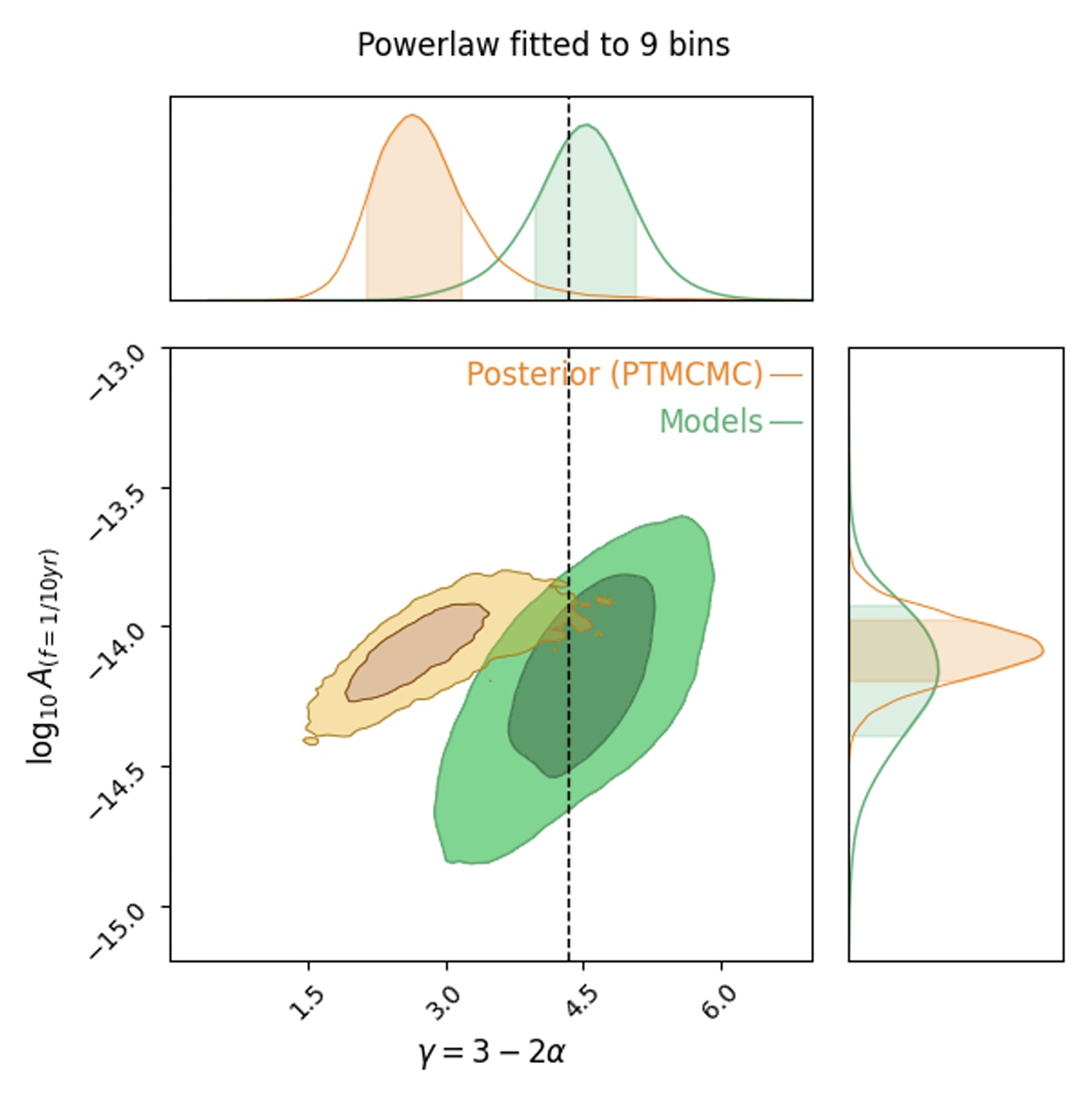}
    \caption{Comparison of a GWB from realizations of SMBHB population models with the constraints on the signal from the NANOGrav 15-year (top) \cite{2023ApJ...952L..37A} and EPTA DR2 (bottom) \cite{2024A&A...685A..94E} data sets. The left column shows the spectra, while the right column shows the power law parameter posteriors.}
    \label{fig:astro}
\end{figure}

\subsection{Astrophysical sources}

The first class of sources of nano-Hertz GWs come from super massive black hole binaries that should be produced by merging galaxies \cite{1978MNRAS.183..341W}. During the initial stage of the galaxy merger, the two black holes will move closer to the center of the gravitational potential well created by the merged galaxies due to dynamical friction \cite{1980Natur.287..307B}. As the SMBHB is slowly formed, interactions with the surrounding environment become more dominant. Energy from the SMBHB orbit can be transferred to either surrounding stars, gaseous disks, another SMBH from a subsequent merger or through numerous other mechanisms \cite{1995ApJ...446..543R,2003ApJ...583..616J,2008MNRAS.390..192S}. When stars come close to the SMBHB without getting absorbed, they can be slingshot out taking energy from the SMBHB, thus shrinking the orbit in a process called stellar hardening. Dust and gas around the SMBHB come also drive the evolution of the binary. When the SMBHB moves through a disk, some particles in the disk can be absorbed, while other particles are accelerated from the gravitational pull of the SMBHB. This process causes the disk to heat up, which carries away energy from the binary. A third type of interaction are triplet SMBH systems, this will be a temporarily state as either two SMBHs merge leaving another binary or the third SMBH is eventually shot out of the triplet system. Many more mechanisms have been proposed, such as capture of small BHs, interactions with dark matter or more exotic particles and so on. All these mechanisms will drive the evolution of the binary. Ultimately, while some binaries may stall, i.e., they do not merge within a Hubble time, some binaries are expected to reach separation well below a parsec, after which the GW emission from the SMBHB will cause the binary to eventually merge within a Hubble time. This is also called the final-parsec problem \cite{2003AIPC..686..201M} and a question that PTA will help to solve. Thus, both a GWB as well as individual SMBHBs emitting GWs are crucial to study the astrophysics of galaxies and their central black holes with a new messenger in GWs.

The GWB as a superposition of the GWs from a population of SMBHBs in the local Universe can be connected and described by galaxy observables and a relation between the galaxy and the BH. The number of galaxy mergers in the Universe should be linked to the number of SMBHBs. It can be modelled from three observables: 1. galaxy stellar mass function describing the density of galaxies, 2. pair fraction describing the number of galaxies expected to merge within a Hubble time and 3. merger time scale. The galaxy merger rate can be derived from these three observables \cite{2019MNRAS.488..401C}. Using a galaxy-black hole relation we can then derive the SMBHB merger rate.
While a relation between the SMBH and its host galaxy has been established, i.e., M-sigma/M-Mbulge relation \cite{2013ARAA..51..511K}, the underlying physics is still to be fully understood.
Detecting GWs will first confirm that SMBHB merge or at least come close enough to emit GWs and second, the strength of the GWB will help to understand the underlying mechanism as well as the galaxy-black hole relation.

The results presented in 2023 have been analyzed assuming that the signal is purely from SMBHBs \cite{2023ApJ...952L..37A, 2024A&A...685A..94E}. Astrophysical SMBHB populations can produce a GWB, which is consistent with the spectrum of the PTA signal. The GWB signal can add information on the parameters on the model and provide better constraints than using electromagnetic observations alone. Both analyses found the merger time and galaxy-black hole relation to be the two dominant observables. As the strength of the GWB can be increased/decreased by having either faster/slower mergers or heavier/lighter black holes, these two observables also link to the final-parsec problem and the galaxy-black hole relation. The large amplitude of the PTA signal is challenging for the astrophysical models, thus, we find a noticable tension between the measured signal and model predictions (see Figure~\ref{fig:astro}). However, this can be solved by extending the parameter space \cite{2023ApJ...952L..37A}. It is also possible that one or multiple strong individual SMBHB sources are present in the Universe \cite{2024A&A...685A..94E}.

An individual loud SMBHB could also be detected by PTAs \cite{2015MNRAS.451.2417R,2021ApJ...915...97X}. Although there are only robust upper limits at the moment, some hints have been seen in the EPTA and NANOGrav, while the PPTA found no hints. However, PTAs are expected to eventually detect individual SMBHBs as the sensitivity improves and more pulsars are timed. Given this, we can use individual SMBHBs as another target for multi-messenger astronomy. Studying an individual system can be advantageous as it proves that SMBHBs exist in the Universe and can be used to study the astrophysics of the host galaxy. Combining information from the GWB and the individual sources can greatly help to improve our model of the full population.
\citep{EPTA+2023a} \citep{rcb+20}
\begin{figure}[!h]
    \centering
    \includegraphics[width=\linewidth]{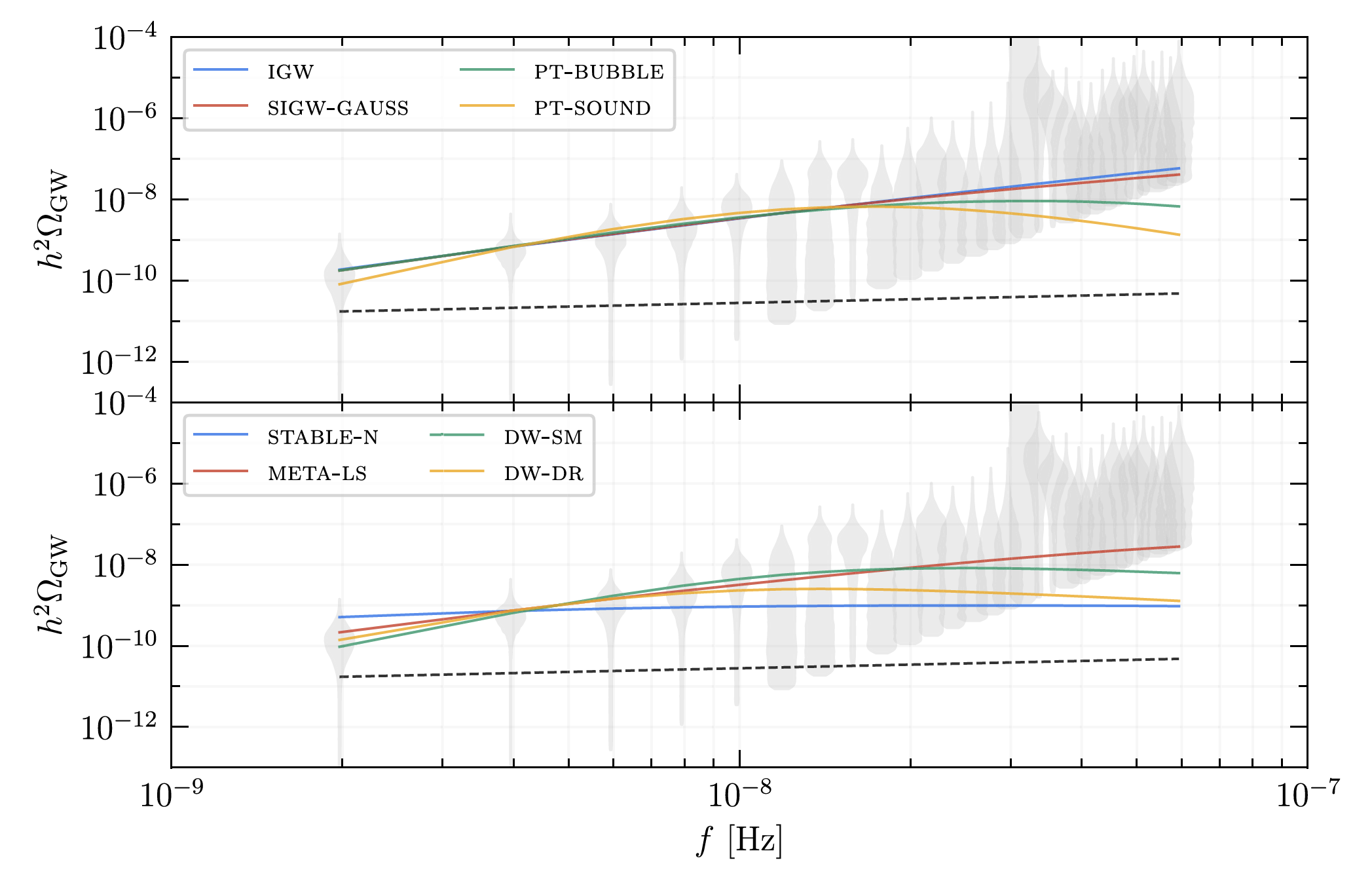}
    \includegraphics[width=0.97\linewidth]{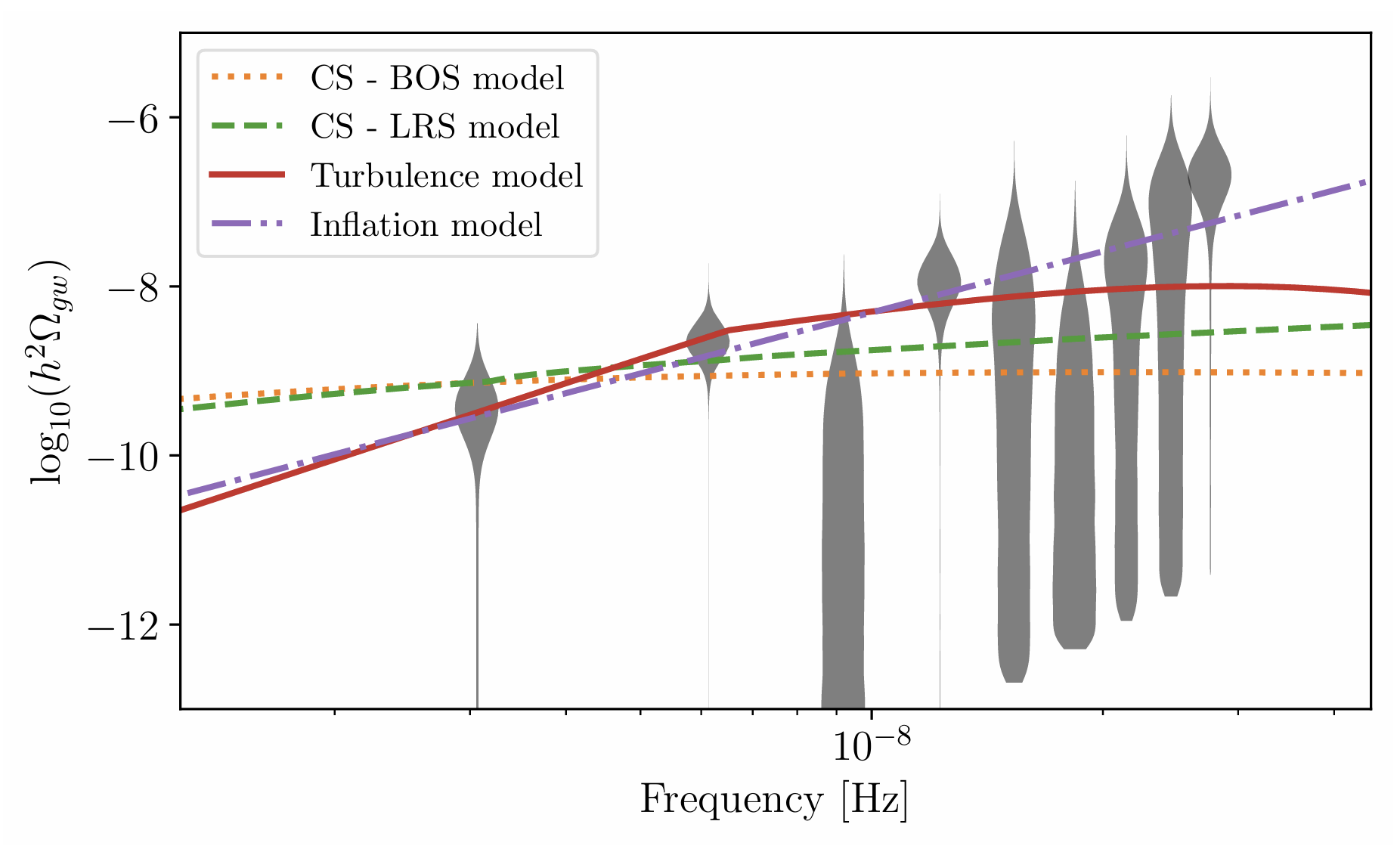}
    \caption{Comparison of a GWB from some selected cosmological and exotic source models with the constraints on the signal from the NANOGrav 15-year (top) \cite{2023ApJ...951L..11A} and EPTA DR2 (bottom) \cite{2024A&A...685A..94E} data sets.}
    \label{fig:new_physics}
\end{figure}

\subsection{Cosmological and exotic sources}

The second class of sources of nanohertz GWs comes from the Early Universe and more exotic sources. These sources are more theoretical in nature, however, a detection would have strong implications on the fundamental physics of the Universe. PTAs provide a unique window into high-energy physics and early-universe processes (e.g., inflation, phase transitions, defects, dark matter). A possible detection and identification of the sources will be crucial for precisely characterizing the origin and implications of these signals. There is a plethora of theoretical models predicting GWs in the nanohertz frequencies, all of which can fit the measured PTA signal very well by constraining the parameters of the model to match the observed spectrum. This review will only be able to capture a small fraction of all possible theories and should only serve to give a general picture from the point of view of the PTA community. A selection of models that have been constrained is shown in Figure \ref{fig:new_physics}.

\subsubsection{Topological defects}

Cosmic strings are one dimensional topological defects whose oscillating loops radiate GWs. PTAs can probe the tension of these with the most recent results setting upper limits at $\log_{10} (G\mu) < -9.77/-10.44$ from the EPTA DR2 for two different models \cite{2024A&A...685A..94E} and $\log_{10} (G\mu) < -9.88$ from the NANOGrav 15-year for a number of models \cite{2023ApJ...951L..11A}.

In two dimensions the topological defects become domain walls, which come from discrete symmetry breaking. The annihilation of these domain walls can produce a flat GW spectrum. Both EPTA and NANOGrav have put constraints on domain walls \cite{2023ApJ...951L..11A,2024A&A...685A..94E}.

\subsubsection{Phase transitions in the early Universe}

Strong first order phase transitions in the early universe (e.g., from dark sectors or QCD) generate GWs via bubble collisions, sound waves, and turbulence. For phase transitions at temperatures $\sim 5–50$ MeV, the sound-wave signal in the PTA frequency range is consistent with the measured spectrum. From $\sim 50-1000$ MeV bubble collisions are able to match the observed PTA signal. A consistent signal is caused by turbulence that peaks around a temperature of $\sim$100 MeV \cite{2023ApJ...951L..11A,2024A&A...685A..94E}.

\subsubsection{Cosmic inflation}

Quantum fluctuations during inflation could generate a stochastic GWB. The signal is generally very weak at nanohertz frequencies, but certain scenarios (like a "blue-tilted" spectrum or "stiff" reheating) could enhance it.
For example, scalar-induced GWs, produced by large curvature perturbations during radiation domination, can produce a significant signal in the PTA band. This signal has been constrained by the EPTA and NANOGrav \cite{2023ApJ...951L..11A,2024A&A...685A..94E}.

\begin{figure}[!h]
    \centering
    \includegraphics[width=\linewidth]{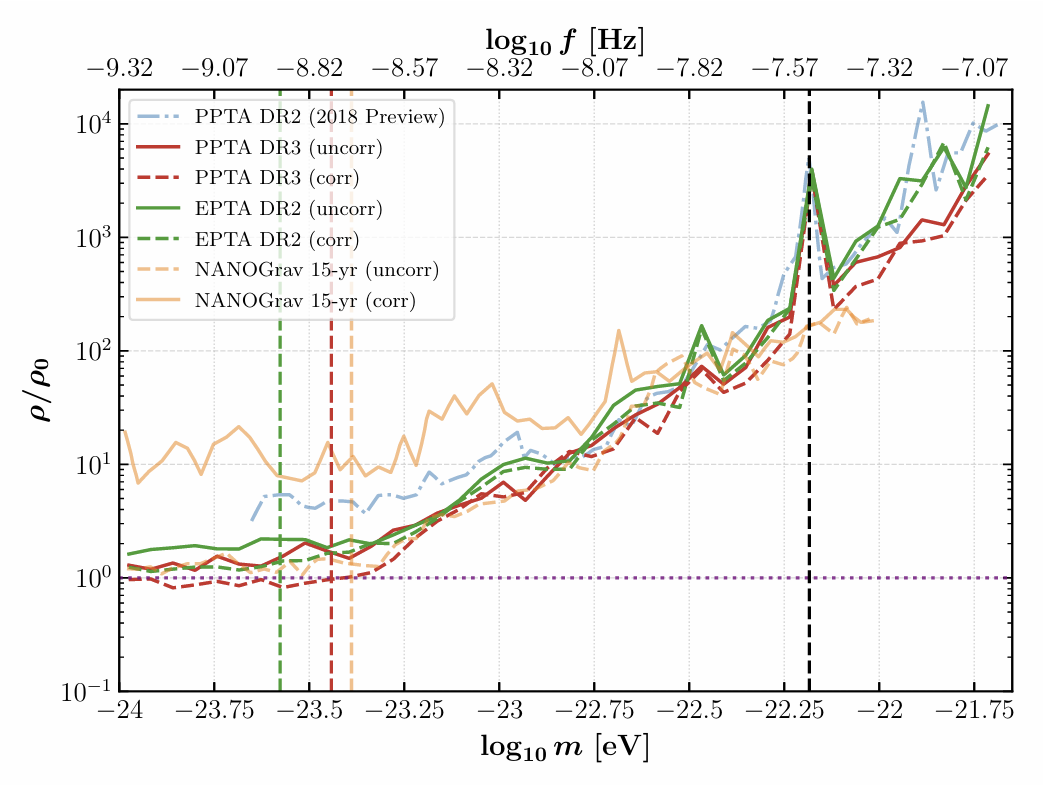}
    \caption{Upper limits on the scalar ULDM density from the PPTA DR3, EPTA DR2 and NANOGrav 15-year data sets \cite{2026_uldm}.}
    \label{fig:uldm}
\end{figure}

\subsubsection{Ultra light dark matter}

The oscillations of ultra light dark matter (ULDM) particles in the galactic potential can induce timing delays in pulsars. PTAs set strong limits on such models. The density of ULDM has been constrained by the data sets from NANOGrav \cite{2023ApJ...951L..11A}, EPTA \cite{2024A&A...685A..94E} and PPTA \cite{2026_uldm}. Since no evidence was found in any of the searches, upper limits were put across the frequency range (equivalent to the ULDM particle mass of $\sim 10^{-24}-10^{-22}$ MeV). Below $\sim$1.6 nHz ($\sim 3 \times 10^{-23}$ MeV) the upper limits approach the measured local dark matter density of $\rho=0.4$ GeV/cm$^3$. Different types of ULDM particles with different couplings have also been constrained by PTAs \cite{2022PhRvD.106h1101W,2023JCAP...09..021W,2026_uldm}. Figure~\ref{fig:uldm} shows the current upper limits on the scalar ULDM density in the Milky Way from the 2023 data sets.

\subsubsection{Other types of sources}

Couplings of axion-like particles to photons could cause cosmic birefringence, rotating pulsar polarization in a correlated way.
Vector fields and Scalar-Tensor couplings can produce narrow-band GWBs or time-dependent potentials detectable by PTAs.
Other Early-Universe GW contributions include, but are not limited to, audible axions, primordial gravitational collapses, relic-neutrino damping effects, modulated reheating and many more. For a more detailed review see \cite{2025OJAp....854243S} and references therein.

\subsection{Deviations from General Relativity}

The underlying assumption of GR can be tested by measuring the properties of the GW signal and comparing them to the predictions from GR. Alternative polarizations, speed of light, mass of graviton and so on are all effects that can impact the expected spatial correlations, i.e., produce deviations from the HD curve. Thus, the majority of tests of GR with PTAs stem from checking how consistent the measured correlations are with the GR predicted HD curve with cosmic uncertainty \citep[e.g.,][]{2023PhRvD.107d3018A}. More details on the current development in the theoretical aspects of PTAs can be found in a companion review \cite{2025IJMPD..3440013B}.

While some violations of GR lead to small deviations from the HD curve, alternative polarizations produce very different spatial correlations. Since current PTA constraints are quite broad, the focus has been to check against theories predicting large deviations from the HD curve. A CRS was detected in the 12.5-year data set from NANOGrav, however, the spatial correlations did not match with HD. Instead, they showed evidence for scalar-transverse polarization \cite{2021SCPMA..6420412C,2021ApJ...923L..22A}. A careful analysis revealed that this spurious signal is likely due to mismodelled noise in a specific pulsar. Removing this pulsar diminishes the significance for the scalar-transverse polarization. This was further strengthened with the 15-year analysis, which shows no indication for alternative polarizations \cite{2024ApJ...964L..14A,2024PhRvD.109h4045C}.

\section{Summary and outlook}
\label{sec:conclu}

In this review, we have summarized the basic concepts underlying PTA detectors for nanohertz GWs, from pulsar observations to GW science. Pulsar timing observations proceed through a sequence of processing steps that ultimately yield pulse TOAs measured at Earth. For each pulsar, a deterministic timing model is constructed to describe the TOAs as a function of time. A noise model is also fitted to capture stochastic processes not described by the deterministic model, within which GW signatures may be present. At the PTA level, one then searches for the characteristic inter-pulsar correlations induced by GWs, which could potentially be of various origins. In this analysis, timing-model parameters are marginalized over, while sampling is performed jointly over GW and pulsar-noise parameters.

The past decade has been particularly fruitful for the PTA community, marked by major milestones on the path toward nanohertz GW detection. Following the most stringent upper limits reported around 2015--2016, a common red signal (CRS) was identified around 2020, and evidence supporting its GW origin emerged in 2023. The nominal signal amplitude, $h_c = 2$–$3 \times 10^{-15}$ at a reference frequency of $1,\mathrm{yr}^{-1}$, is consistent with expectations from a population of supermassive black hole binaries (SMBHBs), as well as with a range of cosmological and more exotic sources. Meanwhile, upper limits on individual CGW sources have steadily improved over the same period. In 2023, a potential hint of a CGW source was reported through its covariance with the GWB, although this result remains inconclusive. In parallel, searches for anisotropies, GW memory signals, and tests of general relativity have become standard components of PTA analyses.

\subsection{The next decade}

PTA science promises to thrive in the years ahead. Currently, the IPTA is actively preparing its next global data release, IPTA Data Release 3 (DR3). This dataset will combine data from established PTAs while incorporating new contributions from emerging PTAs, and will include observations of approximately 130 pulsars spanning over two decades, with frequency coverage extending across roughly two orders of magnitude. The forthcoming IPTA DR3 is therefore expected to represent another significant step forward for GW science with PTAs \citep{2024ApJ...966..105A}, possibly delivering the first conclusive detection of nanohertz gravitational waves. At the same time, the PTA collaborations will continue observations with existing telescopes, progressively extending the baseline of the PTA dataset and adding more PTA pulsars. The use of more powerful signal collection and processing instruments, such as ultra-wideband receivers, will enhance the frequency coverage of PTA data. New data analysis techniques, such as wideband timing, advanced noise models and GPU-based software toolkits, have been actively developed and will be capable of effectively and efficiently reducing larger and more sensitive datasets. Combining the existing data from all IPTA telescopes (and possibly including those from the gamma-ray PTA) ensures a smooth transition into the next era of PTA science. Several years of overlap are necessary to provide a baseline against which observations from the next generation of telescopes can be compared.

The advent of next-generation pulsar timing facilities will open a new era of PTA science, enabling observations at sensitivities and efficiencies unattainable with 100-m-class telescopes. A particularly significant development will be the SKA PTA, which is expected to become a major component of the IPTA effort \citep{sbc+25}. The SKA will not only substantially enhance timing precision but also expand the PTA pulsar sample through its extensive surveys \citep{kgl+25}, which are likely to discover many new MSPs suitable for PTA experiments. Complementing these timing advancements, VLBI astrometry of pulsars using the world's largest radio telescopes, including the SKA, will provide a powerful means of measuring pulsar distances with high accuracy \citep[e.g.,][]{lxl+24}. When determined with sufficient precision, these independent distance measurements can be incorporated directly into gravitational-wave searches, further improving PTA sensitivity by fully exploiting the PTA response to gravitational waves. With a projected sensitivity improvement of a factor of $\gtrsim 3$–4 relative to MeerKAT, the SKA is poised to play a central role in PTA science throughout the 2030s \citep{sbc+25}, particularly in the southern hemisphere. To be constructed on a similar timescale, the Deep Synoptic Array-2000 will also provide an important addition to the overall IPTA sensitivity, especially for the northern hemisphere \citep{mwh+24}. Note that FAST is already delivering sensitivity at the level of next-generation facilities, albeit with limited sky coverage \citep{cpta_dr1}.

With these advancements, PTAs are poised to transition from the current era of emerging GWB signals to robust source characterization and, ultimately, precision gravitational-wave astronomy. Improved sensitivity will enable a decisive detection of the stochastic GWB with high statistical significance, allowing its characteristic spectrum, specifically, the power-law index and any possible low-frequency turnover, to be measured with unprecedented accuracy \citep[e.g.,][]{aaa+24}. Such a well-characterized spectrum will serve as a powerful diagnostic tool, breaking degeneracies between astrophysical supermassive black hole binary (SMBHB) populations and potential cosmological contributions. Beyond the background, PTAs are expected to resolve individual SMBHB systems as continuous gravitational-wave sources, with several confidently detectable sources likely to emerge in the coming decade \citep[e.g.,][]{vvp+24}. These single-source detections will open a unique window for probing the properties and environments of SMBHBs, while also enabling sky localization and, in the most favorable cases, multi-messenger follow-up with electromagnetic observatories. Critically, the combination of a well-measured background and resolved individual sources will allow PTAs to distinguish between competing models of galaxy merger histories and binary evolution channels, marking a major milestone in multi-messenger astrophysics. Extending the PTA data time span will further allow GWs at lower frequencies to be probed, exploring more exotic possible origins of the GWB.

In short, the possible achievable milestones of PTA science over the next decade, in chronological order, may be summarized as follows\footnote{These predictions come with absolutely no guarantee.}: 1. {\bf The first conclusive detection of nanohertz gravitational waves.} With the upcoming IPTA DR3, this could occur within as little as the next few years. 2. {\bf The first detection of a continuous gravitational-wave source.} This is likely to be achieved as current PTA facilities continue operating to improve sensitivity and as data analysis techniques advance over time. 3. {\bf A population study of individual sources.} This will become possible once next-generation PTA telescopes begin operation, raising PTA sensitivity in a way that is particularly beneficial for detecting individual gravitational-wave sources. 4. {\bf Precision characterization of the background.} The accumulation of data from next-generation PTA telescopes will ultimately sharpen the detection significance of the GWB and enable precise measurements of its properties. This will allow detailed analyses of its origin, such as the formation and evolution of supermassive black hole binaries or the processes in the early Universe.

\subsection{The challenges} 

Despite this bright outlook, key challenges persist on both the data acquisition and data analysis fronts. As observations continue with existing facilities and with new instruments coming online, PTAs will increasingly confront issues related to the handling and processing of massive data volumes. Future data processing is likely to request more rigorous and systematic quality control, alongside careful selection procedures for both pulsars and data subsets. Decisions regarding which pulsars to include, and which time spans, frequency coverages, and precision thresholds to adopt, may prove essential for the robust detection significance of the GWB \citep[e.g.,][]{ffs+25}. On the data analysis front, more rigorous and efficient noise models need to be developed and applied to the PTA data sets for the measurement of GW signals. In parallel, development efforts are required to focus on more efficient parameter-space sampling algorithms and the adaptation of PTA analysis pipelines to large-scale parallel computing architectures, such as GPUs.

From the perspective of GW searches, the primary challenges include achieving a decisive detection of the GWB, characterizing the underlying source populations, and leveraging these measurements to probe fundamental physics. A well-measured GWB spectrum can place meaningful constraints on both astrophysical populations and cosmological models. Meanwhile, detecting a large number of individual sources, anisotropies or GW bursts with memory, alongside potential deviations from GR in GW signals, will likely require substantial further improvements in signal quality and mitigation of astrophysical foregrounds and other sources of interference \citep[e.g.,][]{all23}. Once realized, these improvements would enable detailed source classification and fully open the door to multi-messenger astrophysics. Well-constrained levels of anisotropies have the potential to help distinguish between astrophysical and cosmological sources. Crucially, any confirmed deviation from GR would constitute evidence for new physics, potentially heralding a novel frontier in fundamental research.

Finally, looking a bit beyond PTAs, another challenge is to expand the range of frequencies that can be probed by timing pulsars. In particular, pulsar binary systems can be used to probe frequencies both below nanohertz as well as above millihertz \citep{kop97,hmy13,bj22a,bj22b,2026arXiv261107363L}. Better observations are increasing the sensitivity of the binary pulsars towards GWs at all frequencies, at the same time, recent advances have been made to develop novel methods to improve upon the GW upper limits. In the sub-nanohertz regime, \cite{2024PhRvL.132j1403D,2026PhRvD.113b3037Z} propose to use the drift of timing parameters to constrain GWs potentially leading to lower upper limits when applied to real data. Constraining binary pulsar resonances can be used to better constrain GWs at milli- to microhertz frequencies \citep{2022PhRvL.128j1103B,2022PhRvD.105f4021B,2026PhRvL.137i1401F}. Combining all the frequencies from pico- to millihertz will greatly expand the physics we can test with pulsar timing and massively enhance the chances for major breakthroughs in GW astronomy.

\section*{Acknowledgments}

Kuo Liu and Siyuan Chen contribute equally to this article. We thank the anonymous referees for their very constructive suggestions, which helped improve the depth and readability of this paper.

\appendix

\setlength{\bibsep}{-1pt plus -5ex}
\bibliographystyle{ws-ijmpd}

\bibliography{journals,gwrefs,psrrefs-p1,psrrefs-p2,modrefs-p1,modrefs-p2,2024,2025,2026}

\end{document}